%% file: main_neuron_revision.tex
\documentclass{elsarticle}

\usepackage{multirow}
\usepackage{longtable}

\usepackage{lineno}

\biboptions{sort&compress}

\input{my_packages}

\input{my_macros_math}

\input{my_macros_format}

\graphicspath{ {Fig/} }

\begin{document}

\begin{frontmatter}

	
	
	\title{Diffusion MRI simulation of realistic neurons with SpinDoctor and the Neuron Module}
	

	\ead{jingrebecca.li@inria.fr}
	\author[inria1,inria2]{Chengran Fang}
	\author[kth]{Van-Dang Nguyen}
	\author[inria2]{Demian Wassermann}
	\author[inria1]{Jing-Rebecca Li \corref{cor1}}
	\address[inria1]{INRIA Saclay, Equipe DEFI, CMAP, Ecole Polytechnique, 91128 Palaiseau Cedex, France}
	\address[inria2]{INRIA Saclay, Equipe Parietal, 1 Rue Honoré d'Estienne d'Orves, 91120 Palaiseau, France}
	\address[kth]{Department of Computational Science and Technology, KTH Royal Institute of Technology, Sweden}
	
	\cortext[cor1]{Corresponding author}
	
	\begin{abstract}

		The diffusion MRI signal arising from neurons can be numerically simulated by solving the
		Bloch-Torrey partial differential equation.  In this paper we present the Neuron Module that
		we implemented within the Matlab-based diffusion MRI simulation toolbox SpinDoctor.
		SpinDoctor uses finite element discretization and adaptive time integration to solve the Bloch-Torrey partial differential equation for general diffusion-encoding sequences, at multiple b-values and in multiple diffusion directions.		
		In order to facilitate the diffusion MRI simulation of realistic neurons by the research community,
		we constructed finite element meshes for a group of 36 pyramidal neurons and a group of 29 spindle neurons whose morphological descriptions were found in the publicly available neuron repository {\it NeuroMorpho.Org}.
		These finite elements meshes range from having 15163 nodes to 622553 nodes.
		We also broke the neurons into the soma and dendrite branches and created finite elements meshes for
		these cell components. Through the Neuron Module, these neuron and cell components finite element 
		meshes can be seamlessly coupled with the functionalities of SpinDoctor to provide the diffusion MRI signal attributable to spins inside neurons.
		We make these meshes and the source code of the Neuron Module available to the public as an open-source package.
		
		To illustrate some potential uses of the Neuron Module, we show numerical examples of the simulated diffusion MRI signals
		in multiple diffusion directions from whole neurons as well as from the soma and dendrite branches, 
		and include a comparison of the high b-value behavior between dendrite branches and whole neurons. In addition, we demonstrate that the neuron meshes can be used to perform Monte-Carlo diffusion MRI simulations as well. We show that at equivalent accuracy, if only one gradient direction needs to be simulated, SpinDoctor is faster than a GPU implementation of Monte-Carlo, but if many gradient directions need to be simulated, there is a break-even point when the GPU implementation of Monte-Carlo becomes faster than SpinDoctor. Furthermore, we numerically compute the eigenfunctions and the eigenvalues of the Bloch-Torrey and the Laplace operators on the neuron geometries using a finite elements discretization, in order to give guidance in the choice of the space and time discretization parameters for both finite elements and Monte-Carlo approaches. \soutnew{}{Finally, we perform a statistical study on the set of 65 neurons to test some candidate biomakers that can potentially indicate the soma size.  This preliminary study exemplifies the possible research that can be conducted using the Neuron Module.}
		
	\end{abstract}
	
	
\end{frontmatter}
% \linenumbers

% \linenumbers

\section{Introduction}
\label{introduction}

Diffusion magnetic resonance imaging is an imaging modality that can be used
to probe the tissue micro-structure by encoding the incoherent motion of water molecules
with magnetic field gradient pulses \cite{Hahn1950,Stejskal1965,LeBihan1986}.
Using diffusion MRI to get tissue structural information in the mammalian brain has been the focus of much experimental
and modeling work in recent years.

In terms of modeling, the predominant approach up to now has been adding the diffusion MRI signal from simple geometrical components and extracting model parameters of interest.  Numerous biophysical models subdivide the tissue into compartments described by spheres (or ellipsoids), cylinders (or sticks), and the extra-cellular space.  Such modeling work for the brain white matter can be found in \cite{Assaf2008,Alexander2010,zhang2011,Panagiotaki2012}
and for the gray matter in \cite{Jespersen2007,zhang2012,Burcaw2015,Palombo2017a,Palombo2016,Lampinen2017}.
Some model parameters of interest include axon diameter and orientation, neurite density,  dendrite structure,
the volume fraction and size distribution of cylinder and sphere components and the effective diffusion coefficient
or tensor of the extra-cellular space.
More sophisticated mathematical models based on homogenization and perturbations of the intrinsic diffusion coefficient 
can be found in \cite{Novikov2010,Ning2017} and the references contained therein.

Numerical simulations can help deepen the understanding of the relationship between the cellular structure
and the diffusion MRI signal and can play a significant role in the formulation and validation of appropriate models in order
to answer relevant biological questions.  Some recent works that use numerical
simulations of the diffusion MRI signal as a part of model validation include \cite{Jespersen2019,Veraart2019}.
Simulations can be also used to investigate the effect of different pulse sequences and tissue features on the measured signal 
for the purpose of  developing, testing, and optimizing novel MRI pulse sequences \cite{Ianus2016, Drobnjak2011,Mercredi2018,Rensonnet2018}.
In fact, given the recent availability of vastly more advanced computational resources and computer memory,
simulation frameworks have begun to be increasingly used directly as the computational model for tissue parameter estimation  \cite{Palombo2016,Rensonnet2019}.

Two main groups of approaches to the numerical simulation of
diffusion MRI are 1) using random walkers to mimic the diffusion process in a geometrical configuration;
2)  solving the Bloch-Torrey partial differential equation, which describes the evolution of the complex transverse water proton magnetization under the influence of diffusion-encoding magnetic field gradients pulses.

The first group is referred to as Monte-Carlo \soutnew{}{(``MC'' for short)} simulations in the literature and previous works include \cite{Hughes1995, Yeh2013, Hall2009,Palombo2016,Balls2009}.  GPU-based accelerations of Monte-Carlo simulations were proposed in \cite{Nguyen2018a,Waudby2011}.
Some software packages using this approach include
\begin{enumerate}
	\item Camino Diffusion MRI Toolkit, developed at UCL (http://cmic.cs.ucl.ac.uk/camino/);
	\item DIFSIM, developed at UC San Diego (http://csci.ucsd.edu/projects/simulation.html);
	\item Diffusion Microscopist Simulator, \cite{Yeh2013} developed at Neurospin, CEA;
	\item We mention also that the GPU-based Monte-Carlo simulation code described in \cite{Nguyen2018a} is available upon request from the authors.
\end{enumerate}
The works on model formulation and validation for brain tissue diffusion MRI cited above \cite{Jespersen2019,Veraart2019,Ianus2016, Drobnjak2011,Mercredi2018,Rensonnet2018,Palombo2016,Rensonnet2019} all used Monte-Carlo simulations.

The second group of simulations, which up to now has been less often used in diffusion MRI, relies on solving the
Bloch-Torrey partial differential equation (PDE) in a geometrical configuration.  Numerical methods to solve the
Bloch-Torrey equation with arbitrary temporal profiles have been proposed in \cite{Xu2007, Li2014, Nguyen2014, Beltrachini2015}.
The computational domain is discretized either by a Cartesian grid \cite{Xu2007, Russell2012, Li2014} or by finite elements \cite{Hagslatt2003, Loren2005, Moroney2013, Nguyen2014, Beltrachini2015}.
The unstructured mesh of a finite element discretization appeared to be better than a Cartesian grid in both geometry description 
and signal approximation \cite{Nguyen2014}.  For time discretization, both explicit and implicit ODE solvers have been used.
The efficiency of diffusion MRI simulations is also improved by either a high-performance FEM computing framework \cite{Nguyen2016a, Nguyen2018} for large-scale simulations on supercomputers or a discretization on manifolds for thin-layer and thin-tube media \cite{Nguyen2019}.  Finite elements diffusion MRI simulations can be seamlessly integrated with cloud computing resources such as Google Colaboratory notebooks working in a web browser or with the Google Cloud Platform with MPI parallelization \cite{NGUYEN2019106611}.  Our previous works in PDE-based neuron simulations include the simulation of neuronal dendrites using a tree model \citep{Nguyen2015} and (using the techniques we introduce in this work) the demonstration that diffusion MRI signals reflect the cellular organization of cortical gray matter, these signals being sensitive to cell size and the presence of large neurons such as the spindle (von Economo) neurons \cite{Wassermann2018, Menon662601}.

In a recent paper \cite{lid}, we presented a MATLAB Toolbox called SpinDoctor that is a diffusion MRI
simulation pipeline based on solving the Bloch-Torrey PDE using finite elements and an adaptive time
stepping method.  That first version of SpinDoctor focused on the brain white matter.  

It was shown in \cite{lid} that at equivalent accuracy,  SpinDoctor simulations of the extra-cellular space in the white
matter is 100 times faster than the Monte-Carlo based simulations of Camino (http://cmic.cs.ucl.ac.uk/camino/),
and SpinDoctor simulations of a neuronal dendrite tree
is 400 times faster than Camino.
We refer the reader to \cite{lid} for the numerical validation of SpinDoctor simulations with regard to membrane permeability as well as extensions to non-standard pulse sequences and the incorporation of
transverse relaxation.

In this paper, we present a new module of SpinDoctor called the Neuron Module that enables neuron simulations for a group of 36 pyramidal neurons and a group of 29 spindle neurons whose morphological descriptions were found in the publicly available neuron repository {\it NeuroMorpho.Org} \cite{ascoli9247}.
The key to making accurate simulations possible is the use of high quality meshes for the neurons.  For this,
we used licensed software from ANSA-BETA CEA Systems \cite{ansa}
to correct and improve the quality of the geometrical descriptions of the neurons.  After processing, we produced good
quality finite elements meshes for the collection of 65 neurons.  These finite elements meshes range from having 15163 nodes to 622553 nodes.  They are used as input meshes in the Neuron Module, where they can be further refined if required using the built-in option in SpinDoctor.
Currently, the simulations in the Neuron Module enforce homogeneous Neumann boundary conditions,
meaning the spin exchange across the cell membrane is assumed to be negligible.

A recent direction for facilitating Monte-Carlo simulations is the generation of geometrical meshes that aim to
produce ultra-realistic virtual tissues, see \cite{Palombo2019,Ginsburger2019}.
Our work is similar in spirit, with a first step being providing high quality meshes of realistic neurons
for finite elements simulations.  Through the Neuron Module, the neuron finite element meshes can be seamlessly coupled with the functionalities of SpinDoctor to provide the diffusion MRI signal attributable to spins inside neurons for general diffusion-encoding sequences, at multiple diffusion-encoding gradient amplitudes and directions.

We make a note about the software.  The first version of SpinDoctor is a pipeline that constructs surface meshes relevant to the brain white matter and performs diffusion MRI simulations using SpinDoctor's internally constructed meshes.  For technical reasons related to software organization, we decided to put the neuron meshes into a separate and stand-alone pipeline and called it the Neuron Module. The diffusion MRI simulation related routines were copied from the original SpinDoctor pipeline into the Neuron Module pipeline. Within the Neuron Module, we provide additional functionalities that are relevant to treating externally generated meshes.  The Neuron Module and other Modules that we have developed are all grouped under the umbrella of the Matlab Toolbox whose name remains SpinDoctor.  The user is referred to the online User Guide for the technical details of using the Toolbox.  We also mention the existence of another SpinDoctor pipeline called the Matrix Formalism Module \cite{li2019practical} that numerically computes the eigenfunctions and eigenvalues of the Bloch-Torrey and Laplace operators and we will use it to show later in this paper the time and space scales of neuron diffusion MRI simulations.

\section{Theory}

Suppose the user would like to simulate the diffusion
MRI signal due to spins inside a neuron, and assume that the spin exchange across the cell membrane is negligible for the requested simulations.
Let $\Omega$ be the 3 dimensional domain that describes the geometry of the neuron of interest and let
$\Gamma = \partial \Omega$ be the neuron cell membrane.

\subsection{Bloch-Torrey PDE}

\label{PDEsofdiffusionMRI}

In diffusion MRI, a time-varying magnetic field gradient is
applied to the tissue to encode water diffusion.
Denoting the effective
time profile of the diffusion-encoding magnetic field gradient by $f(t)$, and let the
vector $\bg$ contain the amplitude and direction information of the
magnetic field gradient, the complex transverse water proton magnetization
in the rotating frame satisfies the Bloch-Torrey PDE:
\begin{alignat}{4}
	\label{eq:btpde}
	                                         & \frac{\partial}{\partial t}{M(\bx,t)} &                 & = -I\gamma f(t) \bg \cdot \bx \,M(\bx,t)
	+ \nabla \cdot (\Dintr \nabla M(\bx,t)), &                                       & \bx \in \Omega,
\end{alignat}
where $\gamma=2.67513\times 10^8\,\rm rad\,s^{-1}T^{-1}$ is the
gyromagnetic ratio of the water proton, $I$ is the imaginary unit,
$\Dintr$ is the intrinsic diffusion coefficient in the neuron compartment $\Omega$.
The magnetization is a function of position $\bx$ and time $t$,
and depends on the diffusion gradient vector $\bg$ and the time profile $f(t)$.

Some commonly used time profiles (diffusion-encoding sequences) are:
\begin{enumerate}
	\item
	      The pulsed-gradient spin echo (PGSE) \cite{Stejskal1965} sequence,
	      with two rectangular pulses of duration $\delta$, separated by a time
	      interval $\Delta - \delta$, for which the profile $f(t)$ is
	      \be{eq:pgse}
	      f(t) =
	      \begin{cases}
		      1, \quad & t_1 \leq t \leq t_1+\delta,            \\
		      -1,
		      \quad    & t_1+\Delta < t \leq t_1+\Delta+\delta, \\
		      0, \quad & \text{otherwise,}
	      \end{cases}
	      \ee
	      where $t_1$ is the starting time of the first gradient pulse with $t_1
		      + \Delta >T_E/2$, $T_E$ is the echo time at which the signal is measured.
	\item
	      The oscillating gradient spin echo (OGSE) sequence \cite{Callaghan1995,Does2003}
	      was introduced to reach short diffusion times.  An OGSE
	      sequence usually consists of two oscillating pulses of duration
	      $\sigma$, each containing $n$ periods, hence the
	      frequency is $\omega = n\frac{2\pi}{\sigma}$, separated by a time interval
	      $\tau-\sigma$.
	      For a cosine OGSE, the profile $f(t)$ is
	      \be{eq:ogse}
	      f(t) =
	      \begin{cases}
		      \cos{(n \frac{2\pi}{\sigma} t)}, \quad & t_1 < t \leq t_1 + \sigma,           \\
		      -\cos{(n\frac{2\pi}{\sigma} (t-\tau))},
		      \quad                                  & \tau+t_1 < t \leq t_1+\tau + \sigma, \\
		      0, \quad                               & \text{otherwise},
	      \end{cases}
	      \ee
	      where $\tau = T_E/2$.
\end{enumerate}

The PDE needs to be supplemented by interface conditions.  For the neuron simulations within the Neuron Module,
we assume negligible membrane permeability, meaning zero Neumann boundary conditions:
\begin{alignat*}{3}
	\Dintr \nabla M(\bx,t) \cdot \bn & =0,
\end{alignat*}
where $\bn$ is the unit outward pointing normal vector.
The PDE also needs initial conditions:
\begin{align*}
	M(\bx,0) = \rho,\quad
	\label{eq:btpde_initc}
\end{align*}
where $\rho$ is the initial spin density.

The diffusion MRI signal is measured at echo time $t=T_E > \Delta+\delta$ for
PGSE and $T_E > 2\sigma$ for OGSE\@.  This signal is the integral of
$M(\bx, T_E)$:
\be{eq:signal}
S := \int_{\bx\in \bigcup \{\Omega\}} M(\bx, T_E)\;d\bx.
\ee

In a diffusion MRI experiment, the pulse sequence (time profile $f(t)$) is
usually fixed, while $\bg$ is varied in amplitude (and possibly also
in direction).  The signal $S$ is plotted against a quantity called
the b-value.  The b-value depends on $\bg$ and $f(t)$ and is
defined as
\ben
b(\bg) = \gamma^2 \|\bg\|^2 \int_0^{T_E} du\left(\int_0^u f(s) ds\right)^2.
\een
For PGSE, the b-value is \cite{Stejskal1965}:
\be{bvalue_pgse}
b(\bg,\delta,\Delta) = \gamma^2 \|\bg\|^2 \delta^2\left(\Delta-\delta/3\right).
\ee
For the cosine OGSE with {\it integer}\/ number of periods $n$ in each
of the two durations $\sigma$, the corresponding b-value is
\cite{Xu2007}:
\be{bvalue_ogse}
b(\bg,\sigma) =
\gamma^2 \|\bg\|^2 \frac{\sigma^3}{4 n^2\pi^2} = \gamma^2 \|\bg\|^2 \frac{\sigma}{\omega^2}.
\ee
The reason for these definitions is that in a homogeneous medium, the
signal attenuation is $e^{-\Dintr b}$, where $\Dintr$ is the intrinsic diffusion
coefficient.

\section{Method}\label{Method}

\subsection{Constructing finite element meshes of neurons}

In the current version of the Neuron Module, we focus on a group of 36 pyramidal neurons and a group of 29 spindle neurons found in the anterior frontal insula (aFI)
and the anterior cingulate cortex (ACC) of the neocortex of the human brain.
These neurons constitute, respectively, the most common and the largest neuron types in the human brain \cite{nimchinsky1999neuronal, evrard2012economo}.
They share some morphological similarities such as having a single soma and dendrites branching on opposite sides.
The collection of 65 neurons consists of 20 neurons for each type in the aFI, as well as 9 spindles and 16 pyramidals in the ACC.

\begin{figure}[H]
	\centering
	\includegraphics[width=0.4\textwidth]{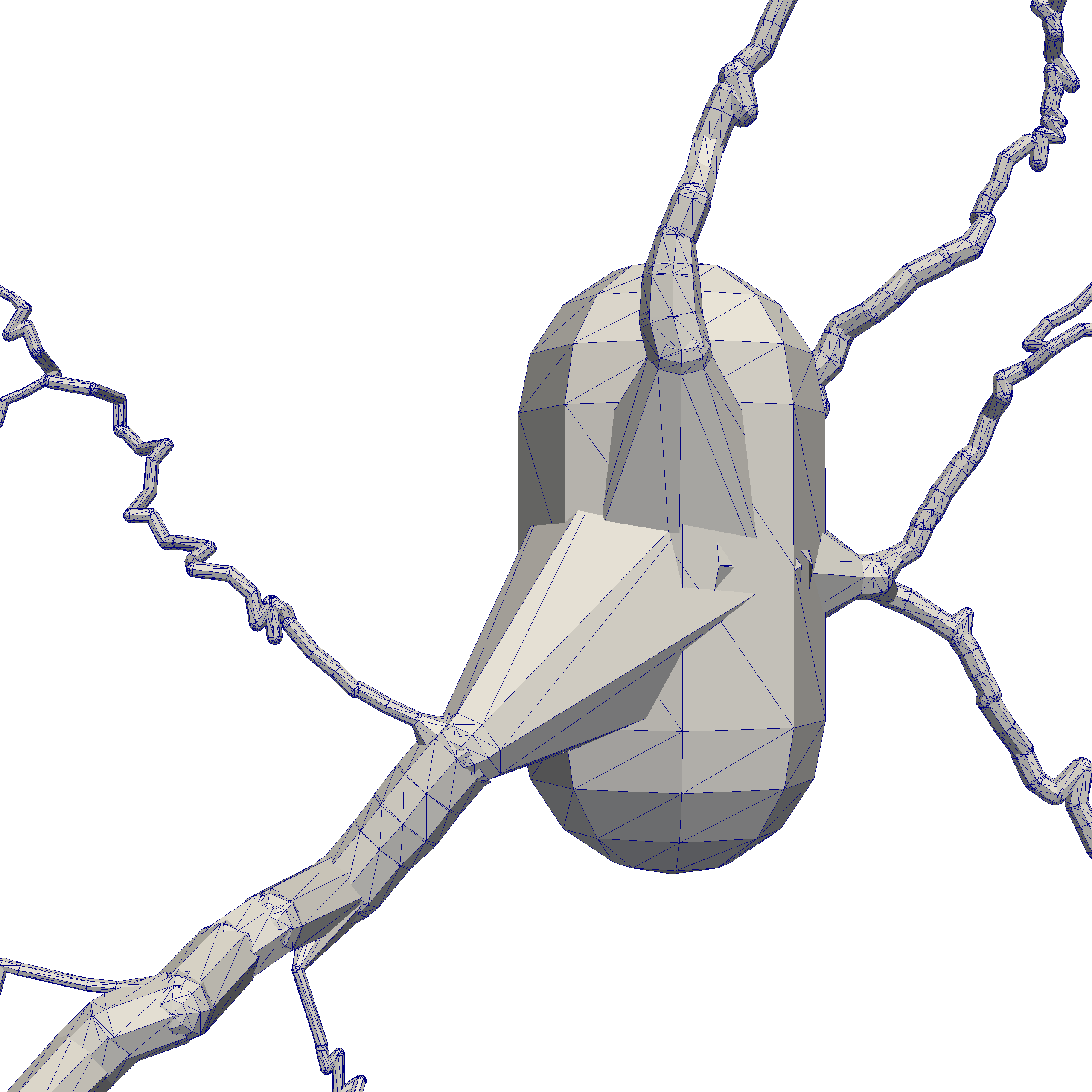}  \quad \
	\includegraphics[width=0.4\textwidth]{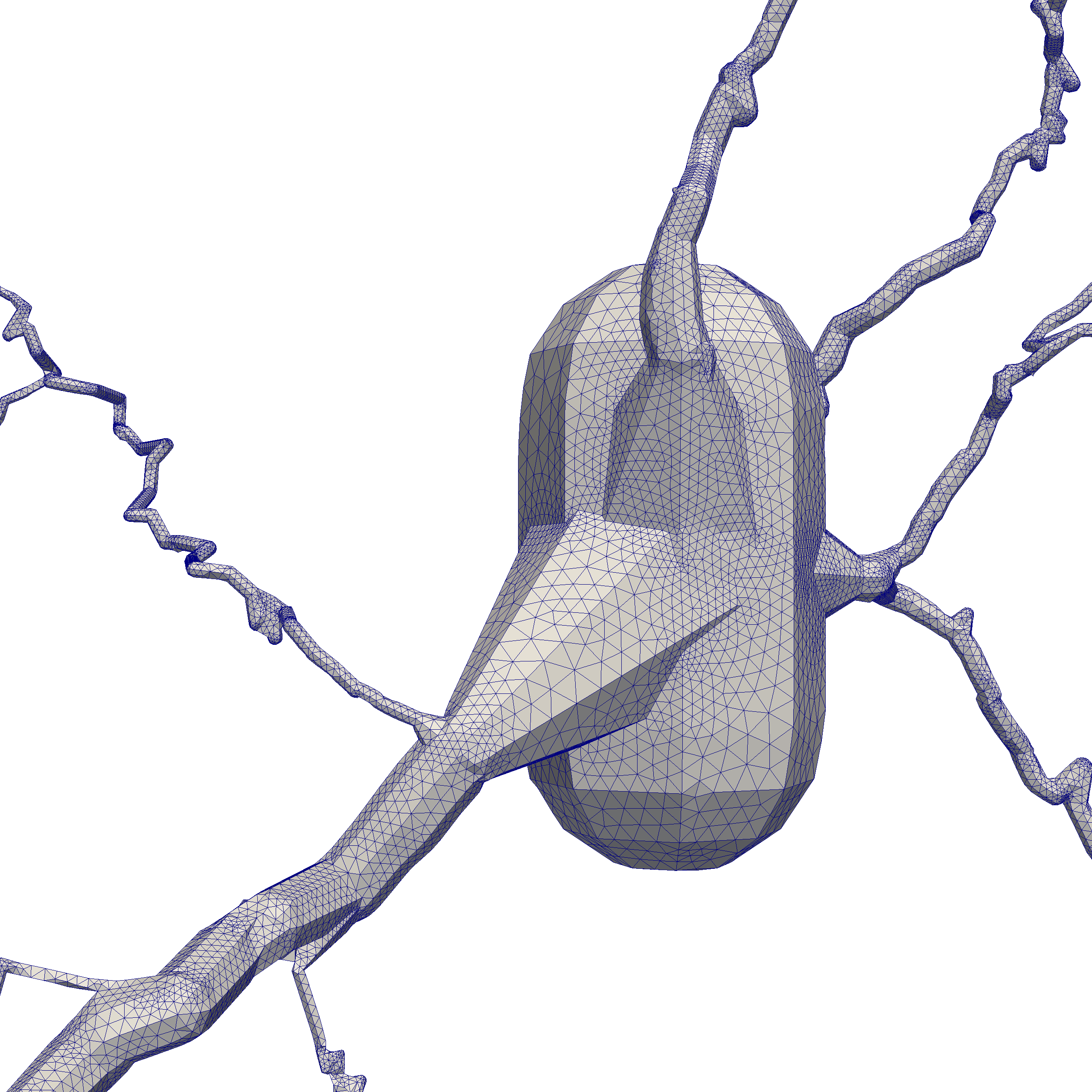}
	\caption{Left: a surface description of a pyramidal neuron published in {\it NeuroMorpho.Org} \cite{ascoli9247}
		which contains
		many self-intersections and proximities;  it cannot be used
		for finite elements mesh generation.
		Right: a new surface triangulation that fixes the self-intersections and proximities; it is ready to be used for
		finite elements mesh generation.}
	\label{fig:02a_pyramidal2aFI_image}
\end{figure}

We started with the morphological reconstructions (SWC files) published in {\it NeuroMorpho.Org} \cite{ascoli9247},
the largest collection of publicly accessible 3D neuronal reconstructions.
These surface descriptions of the neurons cannot be used directly by SpinDoctor
to generate finite elements meshes
since they contain many self-intersections and proximities
(see Figure \ref{fig:02a_pyramidal2aFI_image}, left).
We used licensed software from ANSA-BETA CEA Systems \cite{ansa} to manually correct and improve the quality of the neuron surface descriptions
and produced new surface triangulations (see Figure \ref{fig:02a_pyramidal2aFI_image}, right) that are
ready to be used for
finite elements mesh generation.
The new surface triangulations are passed into the software GMSH \cite{Geuzaine2009} to obtain the
volume tetrahedral meshes.

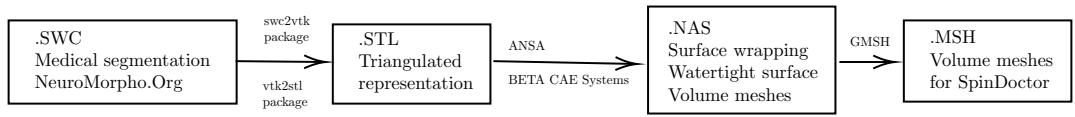
\begin{figure}[H]
	\centering
	\tikzset{every picture/.style={line width=0.75pt}} %set default line width to 0.75pt        
	\begin{center}
		\begin{tikzpicture}[x=0.75pt,y=0.75pt,yscale=-1,xscale=1]
			%uncomment if require: \path (0,235); %set diagram left start at 0, and has height of 235
			%Shape: Rectangle [id:dp966407507826506] 
			\draw  [line width=0.75]  (26.5,85.44) -- (141.5,85.44) -- (141.5,127.44) -- (26.5,127.44) -- cycle ;
			%Shape: Rectangle [id:dp1724648962820413] 
			\draw  [line width=0.75]  (190,87.44) -- (269.5,87.44) -- (269.5,127.44) -- (190,127.44) -- cycle ;
			%Shape: Rectangle [id:dp4463567124637088] 
			\draw  [line width=0.75]  (349,78.44) -- (443.5,78.44) -- (443.5,132.44) -- (349,132.44) -- cycle ;
			%Shape: Rectangle [id:dp06284676406028489] 
			\draw  [line width=0.75]  (478,85.44) -- (562.5,85.44) -- (562.5,126.44) -- (478,126.44) -- cycle ;
			%Straight Lines [id:da4804387084589733] 
			\draw [line width=0.75]    (141,106) -- (184.5,106.42) ;
			\draw [shift={(186.5,106.44)}, rotate = 180.55] [color={rgb, 255:red, 0; green, 0; blue, 0 }  ][line width=0.75]    (10.93,-3.29) .. controls (6.95,-1.4) and (3.31,-0.3) .. (0,0) .. controls (3.31,0.3) and (6.95,1.4) .. (10.93,3.29)   ;
			
			%Straight Lines [id:da73129165103717] 
			\draw [line width=0.75]    (271,106) -- (341.5,107.4) ;
			\draw [shift={(343.5,107.44)}, rotate = 181.14] [color={rgb, 255:red, 0; green, 0; blue, 0 }  ][line width=0.75]    (10.93,-3.29) .. controls (6.95,-1.4) and (3.31,-0.3) .. (0,0) .. controls (3.31,0.3) and (6.95,1.4) .. (10.93,3.29)   ;
			
			%Straight Lines [id:da22457872189044192] 
			\draw [line width=0.75]    (445.5,106.44) -- (471.5,106.44) ;
			\draw [shift={(473.5,106.44)}, rotate = 180] [color={rgb, 255:red, 0; green, 0; blue, 0 }  ][line width=0.75]    (10.93,-3.29) .. controls (6.95,-1.4) and (3.31,-0.3) .. (0,0) .. controls (3.31,0.3) and (6.95,1.4) .. (10.93,3.29)   ;

			% Text Node
			\draw (84,106.22) node [scale=0.7] [align=left] {.SWC\\Medical segmentation\\NeuroMorpho.Org};
			% Text Node
			\draw (232,106.72) node [scale=0.7] [align=left] {.STL\\Triangulated \\representation};
			% Text Node
			\draw (397,105.94) node [scale=0.7] [align=left] {.NAS\\Surface wrapping\\Watertight surface\\Volume meshes};
			% Text Node
			\draw (522.25,105.94) node [scale=0.7] [align=left] {.MSH\\Volume meshes \\for SpinDoctor};
			% Text Node
			\draw (289,98) node [scale=0.5] [align=left] {ANSA };
			% Text Node
			\draw (461,96) node [scale=0.5,color={rgb, 255:red, 0; green, 0; blue, 0 }  ,opacity=1 ] [align=left] {GMSH};
			% Text Node
			\draw (167,90) node [scale=0.5] [align=left] {swc2vtk \\package};
			% Text Node
			\draw (166,123) node [scale=0.5] [align=left] {vtk2stl \\package};
			% Text Node
			\draw (309,116) node [scale=0.5] [align=left] {BETA CAE Systems};
		\end{tikzpicture}
	\end{center}
	\caption{The SWC files from {\it NeuroMorpho.Org} \cite{ascoli9247}  are converted to the STL mesh format by using \lstinline{swc2vtk} \cite{Miyamoto} and \lstinline{vtk2stl}.  Then ANSA was used to generate
	watertight surface and volume meshes of the STL meshes whose output is in the NAS format.  Finally,
	the NAS files were converted to volume tetrahedral meshes in the MSH format by the software GMSH \cite{Geuzaine2009}.}
	\label{fig:pipeline}
\end{figure}

In Figure \ref{fig:pipeline} we summarize the pipeline that takes the SWC format files from {\it NeuroMorpho.Org} \cite{ascoli9247} to the
volume tetrahedral meshes in the MSH format that the users of the Neuron Module will take as the input geometrical
description to the Neuron Module routines that perform diffusion MRI simulations.  This pipeline is provided here for informational purposes, it is not needed to
run diffusion MRI simulations in the Neuron Module.

\begin{figure}[ht!]
	\centering
	\begin{tikzpicture}
		\draw (-2, 0) node[inner sep=0] {\includegraphics[height=8cm]{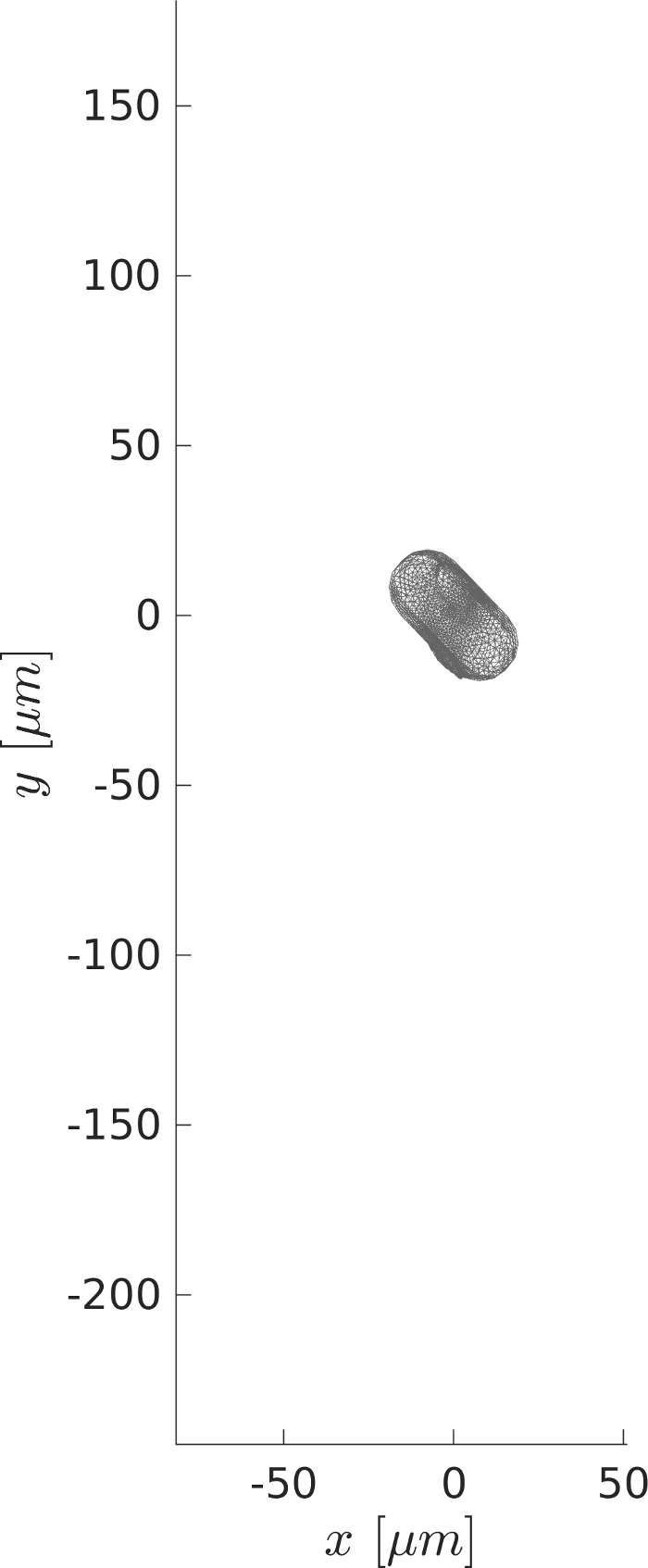}};
		\draw (0.3, 0.5) node {\mbox{\larger[4]$+$}};
		\draw (2.5, 0) node[inner sep=0] {\includegraphics[height=8cm]{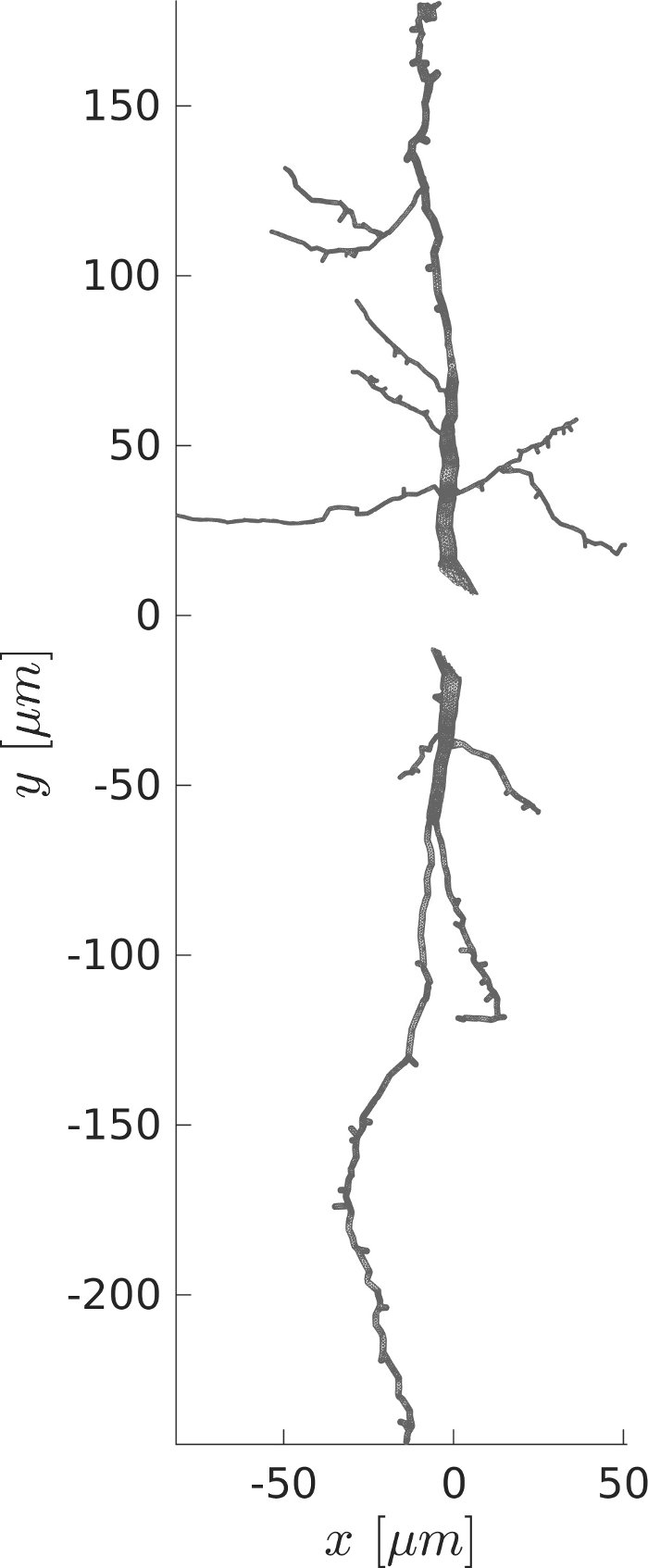}};
		\draw (5.0, 0.5) node {\mbox{\larger[4]$=$}};
		\draw (7, 0) node[inner sep=0] {\includegraphics[height=8cm]{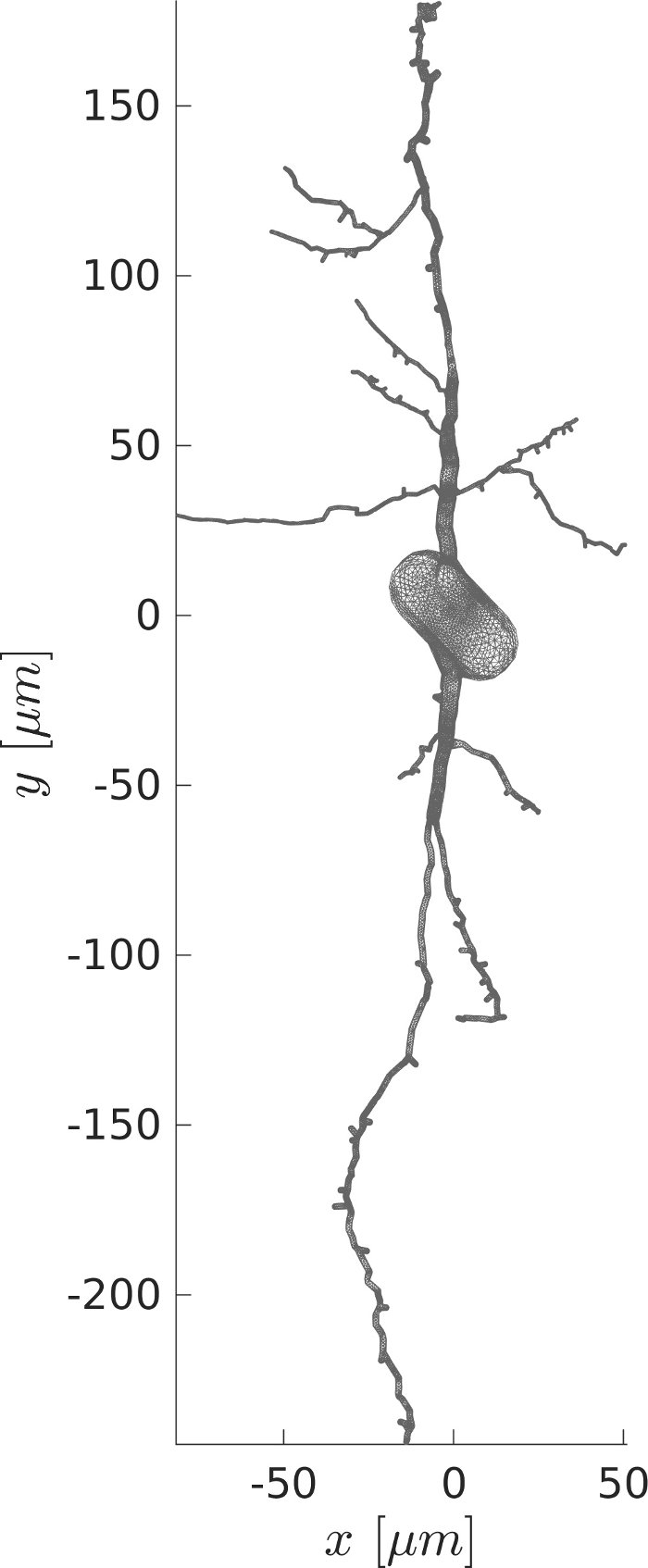}};
	\end{tikzpicture}
	\caption{\label{fig:neuron_broken}  The volume tetrahedral mesh of the spindle neuron {\it 03a\_spindle2aFI} is
		broken into three disconnected geometrical components: the soma and two dendrite branches.}
	\label{fig:03a_spindle2aFI_mesh}
\end{figure}

To further study the diffusion MRI signal of neurons,
we broke the
neurons into disjoint geometrical components: namely,
the soma and the dendrite branches.
We manually rotated the volume tetrahedral mesh of a whole
neuron so that upon visual inspection it lies as much as possible in the $x-y$ plane.  In this orientation,
we cut the volume tetrahedral mesh into sub-meshes of the soma and the dendrite branches.
As an illustration, we show in Figure \ref{fig:neuron_broken} the spindle neuron {\it 03a\_spindle2aFI} volume tetrahedral mesh
broken into sub-meshes of the soma and the two dendrite branches.

\subsection{Diffusion MRI simulations using the Neuron Module}

SpinDoctor \cite{lid} is a MATLAB-based diffusion MRI simulation toolbox.
The first version of SpinDoctor focused on the brain white matter.  It provides the following built-in functionalities:

\begin{enumerate}
	\item  the placement of non-overlapping spherical cells (with an optional nucleus) of different radii close to each other;
	\item  the placement of non-overlapping cylindrical cells (with an optional myelin layer) of different radii close to each other in a canonical configuration where they are parallel to the $z$-axis;
	\item the inclusion of an extra-cellular space that is enclosed either
	      \begin{enumerate}
		      \item in a tight wrapping around the cells; or
		      \item in a rectangular box;
	      \end{enumerate}
	\item  the deformation of the canonical configuration by bending and twisting;
\end{enumerate}
and uses the following methodology:
\begin{enumerate}
	\item it generates a good quality surface triangulation of the user specified geometrical configuration by calling built-in MATLAB computational geometry functions;
	\item it creates a good quality tetrahedron finite elements mesh from the above surface triangulation by calling Tetgen \cite{tetgen}, an external package (the executable files are included in the Toolbox package);
	\item it constructs finite element matrices for linear finite elements on tetrahedra (P1 elements) using routines from \cite{RahmanValdman2013};
	\item it adds additional degrees of freedom on the compartment interfaces to allow permeability conditions for the
	      Bloch-Torrey PDE using the formalism in \cite{Nguyen2014d};
	\item it solves the semi-discretized FEM equations by calling built-in MATLAB routines for solving ordinary differential equations.
\end{enumerate}

The Neuron Module is a stand-alone pipeline that contains all the diffusion MRI simulation routines from the
original SpinDoctor pipeline and adds the relevant routines
to process externally generated neuron meshes.  By neuron meshes we mean watertight surface triangulations and volume tetrahedral meshes.  We do not consider the neuron descriptions from {\it NeuroMorpho.Org} neuron ``meshes".
The Neuron Module takes the neuron watertight surface triangulations or volume tetrahedral
meshes and calls the Neuron Module routines that perform diffusion MRI simulations.  The finite elements mesh package Tetgen \cite{tetgen} contained in the release of SpinDoctor and the Neuron Module is used to refine the input volume tetrahedral meshes, if desired.

The accuracy of the SpinDoctor simulations is tuned using the following three simulation parameters:
\begin{enumerate}
	\item $Htetgen$ controls the finite element mesh size;
	      \begin{enumerate}
		      \item $Htetgen = -1$ means the FE mesh size is determined automatically by the internal algorithm of Tetgen to ensure a good quality mesh (subject to the constraint that the radius to edge ratio of the tetrahedra is no larger than 2.0).
		      \item $Htetgen = h$  requests a desired tetrahedra height of $h$ $\lunit$ (in later versions of Tetgen, this parameter has been changed to the desired volume of the tetrahedra).
	      \end{enumerate}
	\item $rtol$ controls the accuracy of the ODE solve.  It is the relative residual tolerance at all points of the FE mesh at each time step of the
	      ODE solve;
	\item $atol$ controls the accuracy of the ODE solve.  It is the absolute residual tolerance at all points of the FE mesh at each time step of the
	      ODE solve.
\end{enumerate}

\begin{figure}[h]
	\centering
	\includegraphics[width=0.9\textwidth]{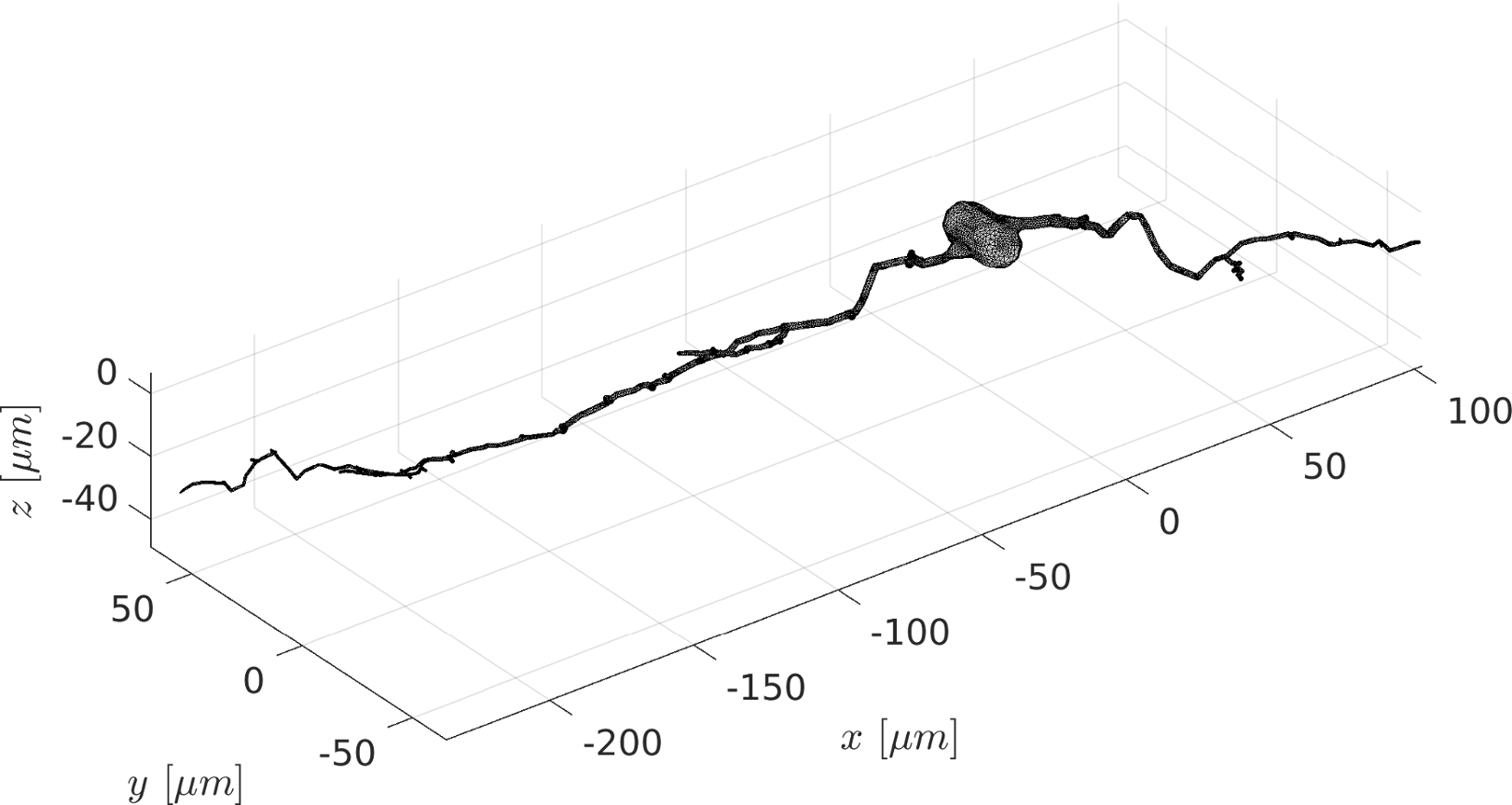}
	\caption{The finite elements mesh (19425 nodes and 60431 elements) of the spindle neuron {\it 03b\_spindle4aACC}.
		\label{fig:03b_spindle4aACC_mesh}}
\end{figure}

All validation, accuracy, and timing simulations were performed on the spindle neuron {\it 03b\_spindle4aACC} (Figure \ref{fig:03b_spindle4aACC_mesh}).
The sizes of the finite elements meshes of this neuron, which are the \textbf{space discretization parameters}, are the following:
\begin{enumerate}[label=Space-\arabic*:, wide=0pt, font=\textbf]
	\item $Htetgen=0.5\mu m$: the finite elements mesh contains 19425 nodes and 60431 elements;
	\item $Htetgen=0.1\mu m$: the finite elements mesh contains 43790 nodes and 157484 elements;
	\item $Htetgen = 0.05\mu m$: the finite elements mesh contains 68536 nodes and 266337 elements;
\end{enumerate}
We ran simulations with the following ODE solver tolerances, which are the \textbf{time discretization parameters}:
\begin{enumerate}[label=Time-\arabic*:, wide=0pt, font=\textbf]
	\item {$rtol = 10^{-2}$, $atol = 10^{-4}$;}
	\item {$rtol = 10^{-3}$, $atol = 10^{-5}$;}
	\item {$rtol = 10^{-4}$, $atol = 10^{-6}$;}
	\item {$rtol = 10^{-6}$, $atol = 10^{-8}$;}
\end{enumerate}

The diffusion MRI experimental parameters, unless otherwise noted, are the following:
\begin{itemize}
	\item the intrinsic diffusion coefficient is $\Dintr = 2 \times 10^{-3}\dunit$;
	\item the diffusion-encoding sequences are PGSE ($\delta = 2.5\tunit$, $\Delta = 5\tunit$), PGSE ($\delta = 10\tunit$, $\Delta = 43\tunit$), PGSE ($\delta = 10\tunit$, $\Delta = 433\tunit$);
	\item the b-values are $b =\{1000,4000\}\bunit$;
	\item 10 or 90 gradient directions were simulated, uniformly distributed on the unit semi-circle in the $x-y$ plane.
\end{itemize}

\section{Numerical results}

\label{sec:numericalresults}

\subsection{Validation of simulations}
In this section, we validate our simulations by refining in space (making the finite elements smaller) and refining in time (decreasing the error tolerances of the ODE solver).
We took the simulation with the finest space discretization ({\textbf{Space-3}}: $Htetgen = 0.05\mu m$) and the finest time discretization
({\textbf{Time-4}}: $rtol = 10^{-6}$, $atol = 10^{-8}$) as the reference solution.

\begin{figure}[H]
	\centering
	\includegraphics[width=0.9\textwidth]{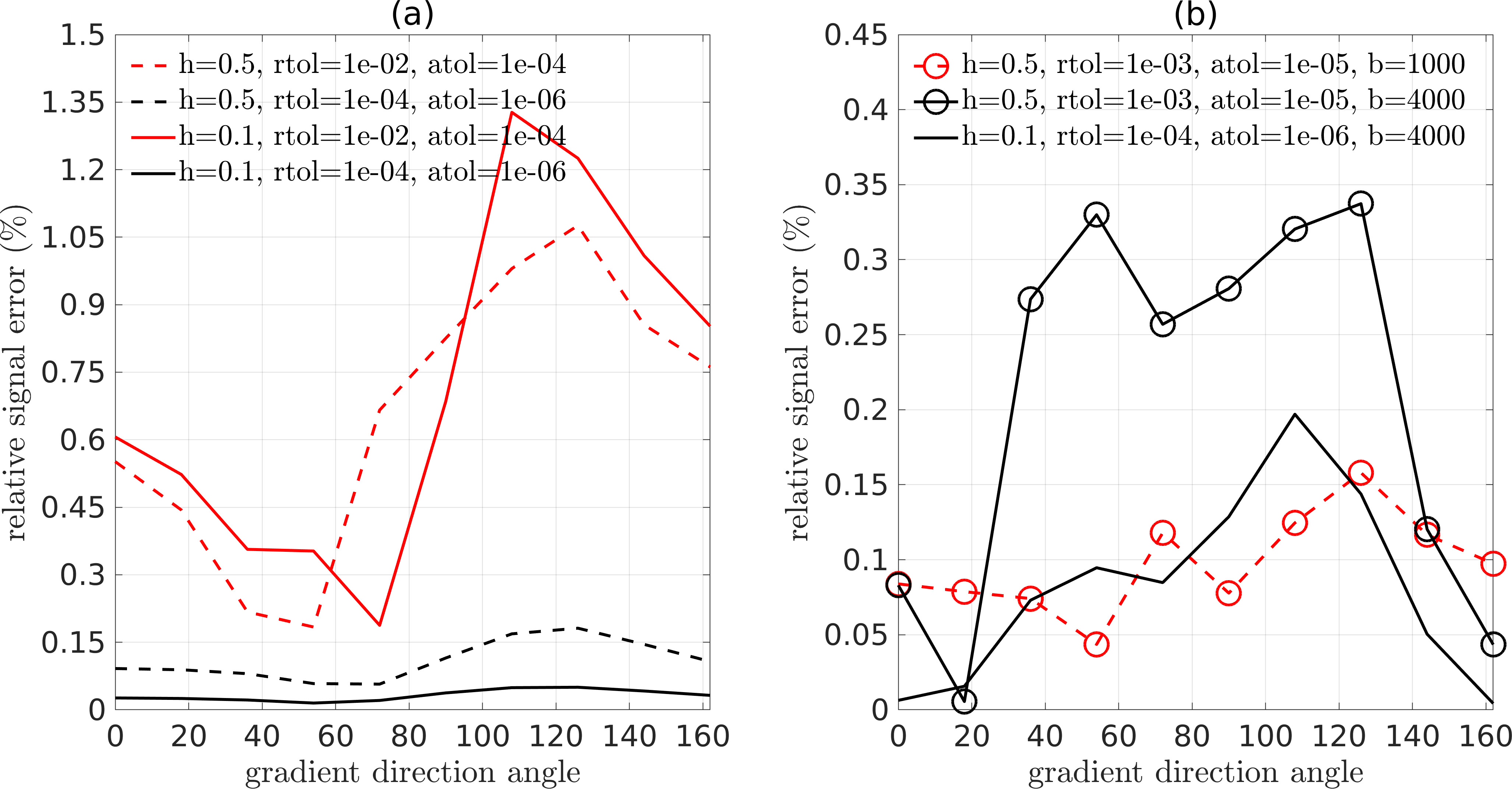}
	\caption{The relative errors between the reference signal and the simulated signals for the neuron {\it 03b\_spindle4aACC}.
		10 gradient directions uniformly placed on the unit semi-circle in the $x-y$ plane were simulated.  
		The gradient direction angle is given with respect to the $x$-axis.
		The simulations with large relative errors are discarded for the clarity of the plots.  
		The diffusion coefficient is $2 \times 10^{-3}\dunit$.  
		The gradient sequence is PGSE ($\delta=10\tunit$, $\Delta=43\tunit$).  In the legend, $h$ denotes $Htetgen$. 
		(a) $b=1000\bunit$. (b) $b=1000\bunit$ and $b=4000\bunit$.
		\label{fig:SD_validation}
	}
\end{figure}

In Figure \ref{fig:SD_validation}, we show the relative signal errors (in percent) between several simulations
and the reference solution:
% \be{}
% E(b) = \frac{\left |S^{Ref}(b)-S^{Simul}(b)\right |}{S^{Ref}(b==0)}\times 100.
% \ee
\be{}
E = \frac{\left |S^{Ref}-S^{Simul}\right |}{S^{Ref}}\times 100,
\label{eq:signal_error}
\ee
for PGSE ($\delta = 10\tunit$, $\Delta = 43\tunit$).
By looking at the difference between the reference solution and the coarser SpinDoctor simulations in Figure \ref{fig:SD_validation}, we estimate that the accuracy of the reference solution is within 0.05\% of the true solution 
at $b=1000\bunit$ and it is within 0.2\% of the true solution at $b=4000\bunit$.
Figure \ref{fig:SD_validation}(a) shows that for $b=1000\bunit$, by refining the time discretization
from \textbf{Time-1} to \textbf{Time-3}, the maximum $E$ over 10 gradient directions went from around 1.06\%
down to 0.2\% for space discretization \textbf{Space-1} and from 1.35\% to 0.05\% for space discretization \textbf{Space-2}.  We see in Figure \ref{fig:SD_validation}(b) that by using
\textbf{Space-1} and \textbf{Time-2}:
\be{}
Htetgen = 0.5\lunit, rtol = 10^{-3}, atol = 10^{-5}
\label{eq:simulation_parameters}
\ee
the maximum $E$ over 10 gradient directions is less than
$0.2\%$ for $b=1000\bunit$ and less than $0.35\%$ for $b=4000\bunit$.
We will choose the above set of simulation parameters in Eq.\ref{eq:simulation_parameters} for the later simulations, unless otherwise noted.

\subsection{Diffusion directions distributed in two dimensions}

We generated 90 diffusion directions uniformly distributed on the
unit semi-circle lying in the $x-y$ plane (plotting
180 directions on the unit circle due to the symmetry of $\bg$ and 
$-\bg$) and computed the diffusion MRI signals in these 180 directions for three sequences:
\begin{itemize}
	\item PGSE ($\delta = 2.5\tunit$, $\Delta = 5\tunit$);
	\item PGSE ($\delta = 10\tunit$, $\Delta = 43\tunit$);
	\item PGSE ($\delta = 10\tunit$, $\Delta = 433\tunit$);
\end{itemize}
The simulation parameters are specified in Eq.\ref{eq:simulation_parameters}.  With this choice, we verified
that the signal is within 1\% of the reference solution for all geometries
(the whole neuron, the soma, the two dendrites branches) for the three gradient sequences simulated.

\begin{figure}[H]
	\centering
	\includegraphics[width=0.9\textwidth]{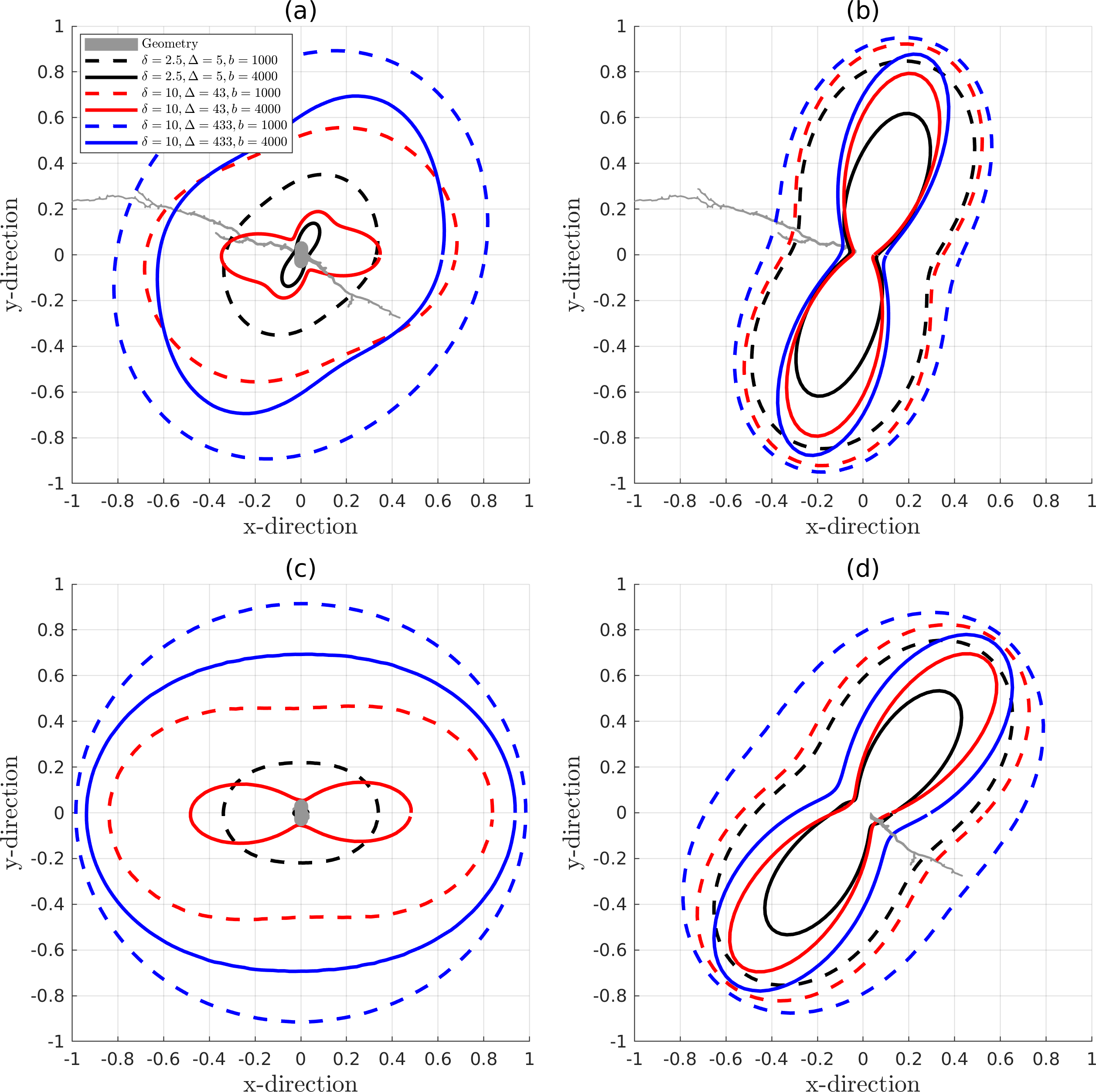}
	\caption{The diffusion MRI signals in 180 directions lying on the $x-y$ plane, uniformly distributed on the unit circle. The distance from each data point to the origin represents the magnitude of the normalized signal which is dimensionless.  The simulation parameters are $rtol = 10^{-3}, atol = 10^{-5}, Htetgen = 0.5\lunit$.  The diffusion coefficient is $2 \times 10^{-3}\dunit$. (a) the whole neuron (finite elements mesh: 19425 nodes and 60431 elements). (b) the dendrite1 (finite elements mesh: 10825 nodes and 30617 elements). (c) the soma (finite elements mesh: 5842 nodes and 26160 elements). (d) the dendrite2 (finite elements mesh: 6444 nodes and 24051 elements).
		\label{fig:03b_spindle4aACC_2D_HARDI}}
\end{figure}

The results for the spindle neuron {\it 03b\_spindle4aACC} are
shown in Figure \ref{fig:03b_spindle4aACC_2D_HARDI}. We plot the normalized signals:
\ben
\left |\frac{S}{S(b=0)}\right |
\een
in the 180 diffusion directions in the $x-y$ plane. The finite elements meshes of the geometries simulated are superimposed on the plots for a better visualization.

It can be seen that the dendrite branch diffusion signal shape
is more like an ellipse at $b=1000\bunit$, whereas at $b=4000\bunit$
the shape is non-convex.  The signal shape of the soma is like an
ellipse except for $b=4000\bunit$ at the two shorter diffusion times.  At the two shorter
diffusion times, the soma signal magnitude at $b=4000\bunit$ is much
reduced with respect to the magnitude at $b=1000\bunit$,
in contrast to the dendrite branches, where the difference in the
signal magnitude between the two b-values is not nearly as significant.
For the soma, at the long diffusion time, there is not the large reduction in the signal magnitude 
between $b=1000\bunit$ and $b=4000\bunit$.

By visual inspection, at the lower b-value, the signal in the whole neuron is close to the volume weighted sum of the signals from the three 
cell components (the soma, the upper dendrite branch and the lower dendrite branch).  A quantitative study is conducted in section \ref{sec:exchange_between_soma_dendrites}.

\subsection{Exchange effects between soma and dendrites}
\label{sec:exchange_between_soma_dendrites}
\begin{figure}[H]
	\centering
	\includegraphics[width=1\textwidth]{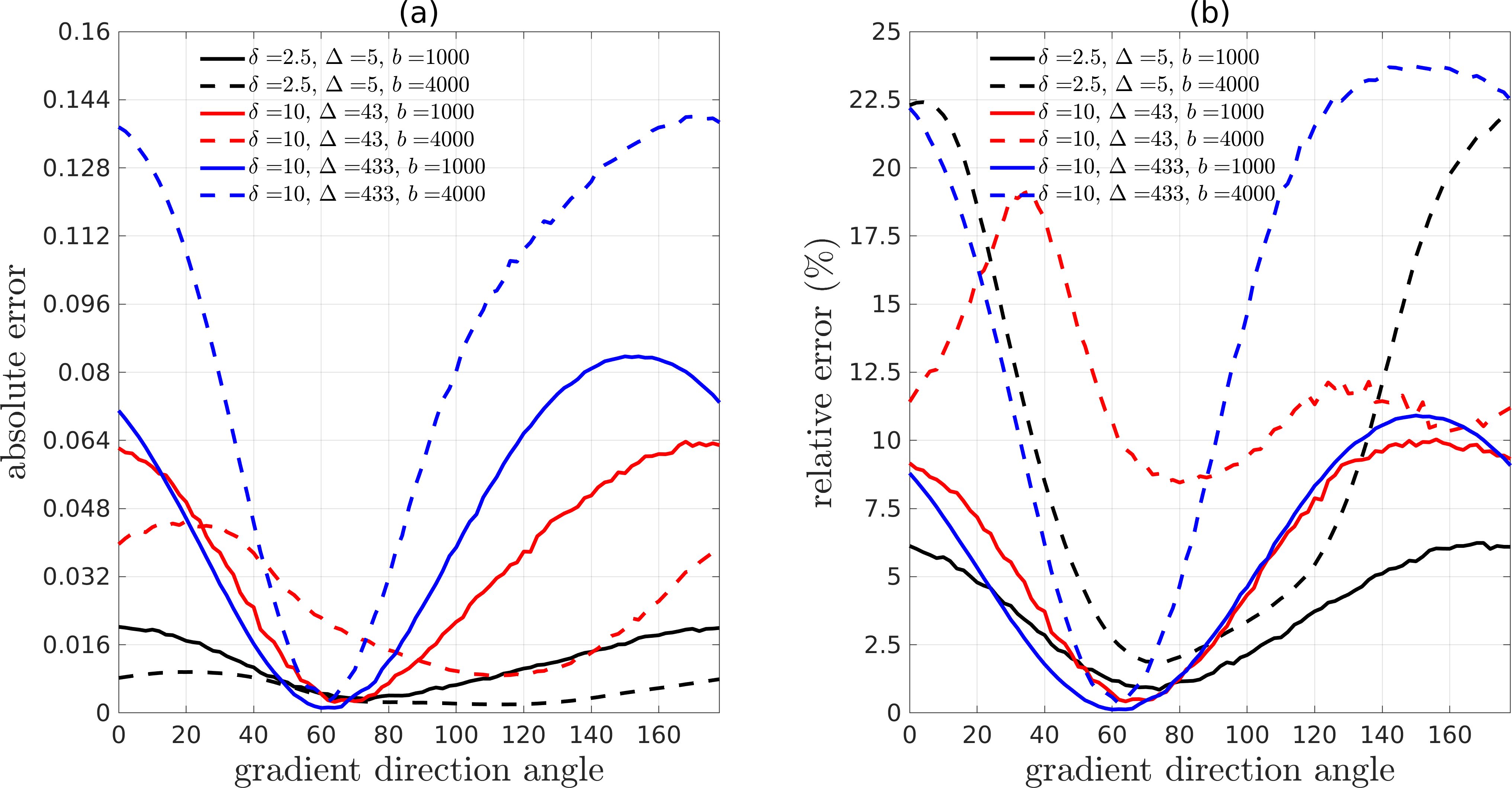}
	\caption{(a) The absolute error between volume weighted
		composite signal and whole neuron signal. (b) The relative error between volume weighted composite
		signal and whole neuron signal.  90 gradient directions uniformly placed on the unit semi-circle in the $x-y$ plane were simulated.  
		The gradient direction angle is given with respect to the $x$-axis.  The position of the neuron 
		can be seen in Figure \ref{fig:03b_spindle4aACC_2D_HARDI}.\label{fig:error_compositesignal}}
\end{figure}

Here we compare the volume weighted composite signal of the 3 cell parts
\be{}
S_{composite} = \frac{V_{soma}S_{soma}+V_{dendrite1}S_{dendrite1}+V_{dendrite2}S_{dendrite2}}{V_{neuron}}
\ee
and compare it to the signal of the whole neuron in the different gradient directions.
In Figure \ref{fig:error_compositesignal} we see that the signal difference between the two is larger at longer diffusion times and at higher b-values. The error also presents a gradient-direction dependence.  According to Figure \ref{fig:03b_spindle4aACC_2D_HARDI} and Figure \ref{fig:error_compositesignal}, we can see that the error is larger in the direction parallel to the longitudinal axis of the neuron than in the direction perpendicular to the longitudinal axis.

\subsection{High b-value behavior}
\label{sec:simul_highb}

\begin{figure}[H]
	\centering
	\includegraphics[width=0.49\textwidth]{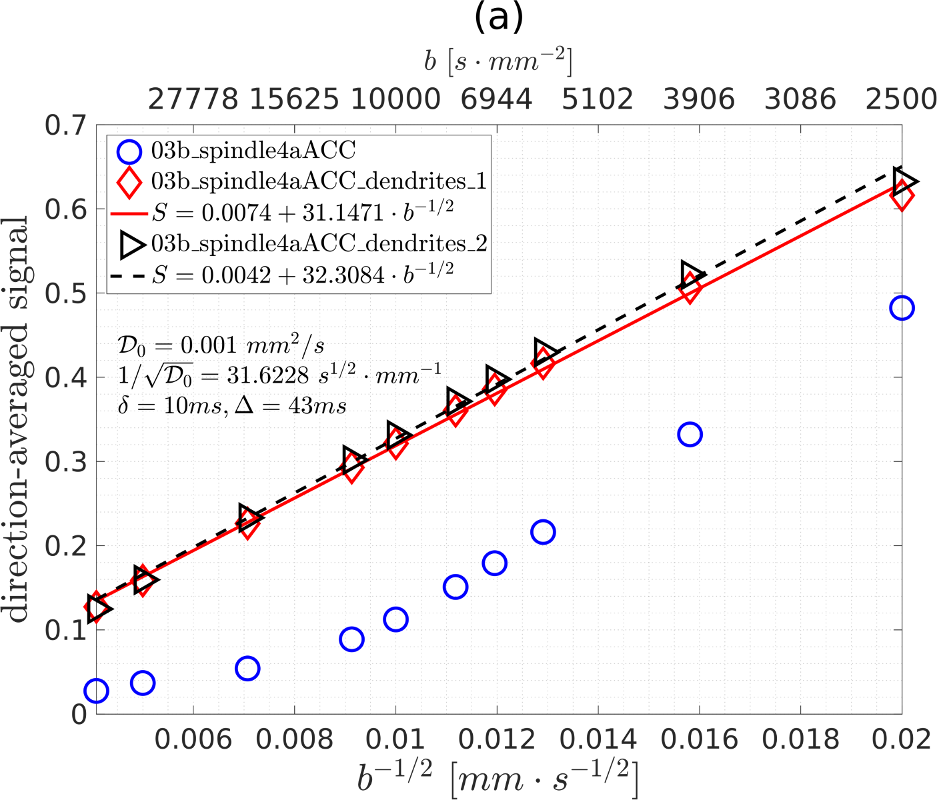}
	\includegraphics[width=0.49\textwidth]{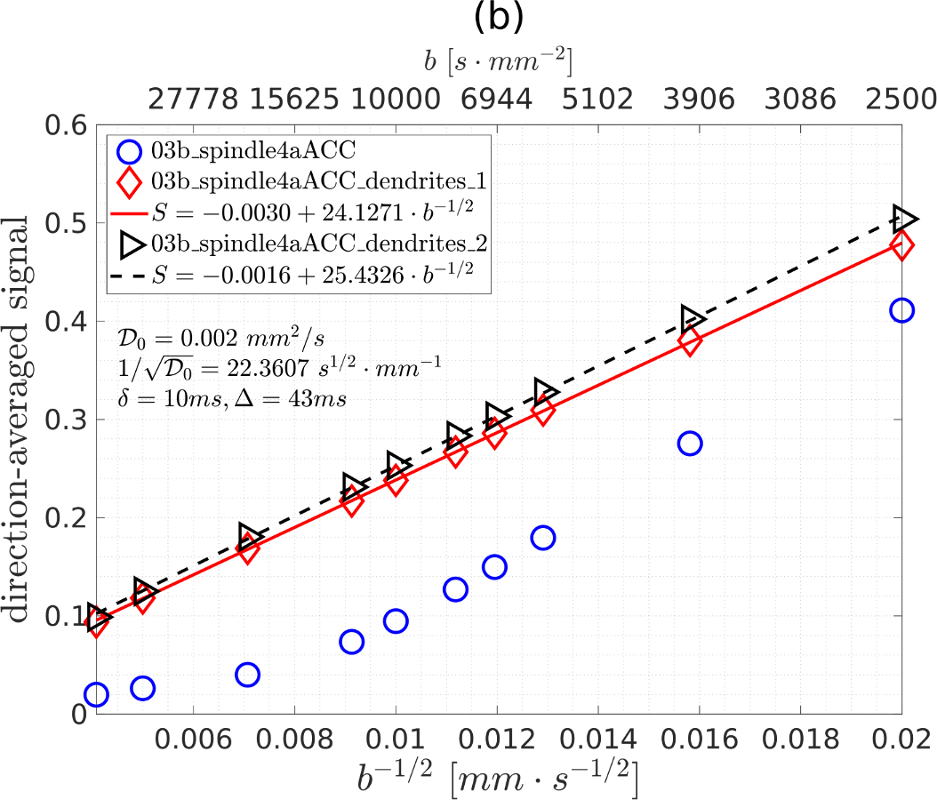}
	\caption{The direction-averaged signal for the neuron {\it 03b\_spindle4aACC}.
		The $S_{ave}(b)$ is averaged over 120 diffusion directions, uniformly distributed in the unit sphere, 
		and it is normalized so that $S_{ave}(b=0)=1$.  The simulation parameters are $rtol = 10^{-3}, atol = 10^{-5}, Htetgen = 0.5\mu m$. The diffusion-encoding sequence is PGSE ($\delta = 10\tunit$, $\Delta = 43\tunit$). 
		The b-values are $b=\{60000, 40000, 20000, 12000, 10000, 8000, 7000, 6000, 4000, 2500\} \bunit$.
		(a) $\Dintr = 1 \times 10^{-3}\dunit$.  (b) $\Dintr = 2 \times 10^{-3}\dunit$.}
	\label{fig:high_bvalue_fitting}
\end{figure}

In \cite{Veraart2019} it was shown experimentally that the diffusion MRI signal of
tubular structures such as axons exhibits a certain high $b$-value behavior,
namely, the diffusion direction averaged signal, $S_{ave}(b)$, is linear in $\frac{1}{\sqrt{b}}$ at high $b$-values:
\be{}
\label{eq:power_law}
S_{ave}(b)  \equiv \int_{\|\bug\| = 1} S_\bug (b) d\bug \sim c_0 + c_1 \frac{1}{\sqrt{b}}.
\ee
Because the dendrites of neurons also have a tubular structure, we test whether
the diffusion direction averaged signal, $S_{ave}(b)$, of dendrite branches also exhibits 
the above high $b$-value behavior.  We computed $S_{ave}(b)$ for the whole neuron as well as its two dendrite branches, 
averaged over 120 gradient directions uniformly distributed
in the unit sphere.  The results are shown in Figure \ref{fig:high_bvalue_fitting}.  We see clearly the
linear relationship between $S_{ave}(b)$ and $\frac{1}{\sqrt{b}}$
in the dendrite branches for $b$-values in the range $2500\bunit \leq b \leq 20000\bunit$.
In contrast, in the whole neuron, due to the presence of the soma, such linear relationship is not exhibited.
By simulating for both $\Dintr = 2\times 10^{-3} \dunit$ and $\Dintr = 1\times 10^{-3} \dunit$
we see that the fitted slope $c_1$ is indeed $\frac{1}{\sqrt{\Dintr}}$.

\subsection{GPU Monte-Carlo simulation on neuron meshes}
\label{sec:simul_gpumc}

We show that one can also use the neuron meshes we provide with Monte-Carlo diffusion MRI simulations.
In particular, we ran the GPU implementation of Monte-Carlo simulations described in \cite{Nguyen2018a},
which provides two surface representation methods, 
namely the octree method and the binary maker method.  Since we only have access to the GPU Monte-Carlo implementation with the 
octree method and the binary maker is an unconventional representation method, the octree method is used for the simulations.

The Monte-Carlo equivalent of the \textbf{space discretization parameter} is the number of spins placed
in the geometry and we simulated with the following choices:
\begin{enumerate}[label=MC Space-\arabic*:, wide=0pt, font=\textbf]
	\item $5\times 10^{5}$ spins;
	\item $1\times 10^{6}$ spins;
	\item $2 \times 10^{6}$ spins;
\end{enumerate}
The Monte-Carlo equivalent of the \textbf{time discretization parameter} is the time step size and we
simulated with the following choices:
\begin{enumerate}[label=MC Time-\arabic*:, wide=0pt, font=\textbf]
	\item $dt = 0.1\tunit$;
	\item $dt = 0.01\tunit$;
	\item $dt=0.005\tunit$;
	\item $dt = 0.001 \tunit$;
\end{enumerate}

To compare with an available analytical solution, we computed the Matrix Formalism\cite{Callaghan1997,Barzykin1999}  signal of a rectangle cuboid of size $L_x \times L_y \times L_z$ using the 
analytical eigenfunctions and eigenvalues of the rectangle cuboid\cite{Grebenkov2007,Ozarslan2009,Drobnjak2011a,Grebenkov2010a}.  
We chose $L_x = 3\lunit$, $L_y = 100 \lunit$, $L_z=1\lunit$ to be close to the size of the dendrite branches.  We computed the diffusion MRI signal in 10 gradient directions uniformly placed on the unit semi-circle in the $x-y$ plane. 
We define the maximum relative error to be:
\be{eq:emax} E_{max} = \max_{\text{10 directions in x-y plane}} \frac{\left |S^{Ref}-S^{Simul}\right |}{S^{Ref}}\times 100,
\ee
where the reference solution is the analytical signal from Matrix Formalism.
In Table \ref{table:accuracy_rectangle} we see that there is a straightforward reduction of the error in the SpinDoctor 
simulations as we decrease the ODE solver tolerances and decrease $Htetgen$,
the $E_{max}$ going from 0.33\% to 0.02\% for $b=1000\bunit$
and the $E_{max}$ going from 0.59\% to 0.03\% for $b=4000\bunit$.  However, for GPU Monte-Carlo, 
the reduction of $E_{max}$ is not consistent as the number 
of spins is increased and $dt$ is reduced.  This means that we cannot use the finest GPU Monte-Carlo simulation as a reference
solution because it is not guaranteed that this simulation is the most accurate one.  
Nevertheless, we see GPU Monte-Carlo has an $E_{max}$
ranging from $0.15\%$ to $0.67\%$ for $b=1000\bunit$ and $E_{max}$ ranges from $0.95\%$ to $6.11\%$ for $b=4000\bunit$,
compared to the reference analytical Matrix Formalism signal.
\begin{table}[H]
	\centering
	\begin{tabular}{|c|c|c|c|c|c|}
		\hline
		\multirow{3}{*}{\begin{tabular}[c]{@{}c@{}}SpinDoctor\\ $\delta=10ms$, $\Delta=43ms$\end{tabular}} & \multicolumn{2}{c|}{\multirow{2}{*}{$E_{max}$}} & \multirow{3}{*}{\begin{tabular}[c]{@{}c@{}}GPU Monte-Carlo\\ $\delta=10ms$, $\Delta=43ms$\end{tabular}} & \multicolumn{2}{c|}{\multirow{2}{*}{$E_{max}$}}                       \\
		                                            & \multicolumn{2}{c|}{}                           &                                             & \multicolumn{2}{c|}{}                                                 \\ \cline{2-3} \cline{5-6}
		                                            & $b=1000$                                        & $b=4000$                                    &                                                 & $b=1000$ & $b=4000$ \\ \hline
		\begin{tabular}[c]{@{}c@{}}$h=0.5\mu m$\\ $rtol=10^{-3}$, $atol=10^{-5}$\end{tabular}                  & 0.33                                            & 0.59                                        & \begin{tabular}[c]{@{}c@{}}$5\times 10^5$spins\\ $dt=0.01ms$\end{tabular}                      & 0.15     & 1.79     \\ \hline
		\begin{tabular}[c]{@{}c@{}}$h=0.5\mu m$\\ $rtol=10^{-4}$, $atol=10^{-6}$\end{tabular}                  & 0.16                                            & 0.39                                        & \begin{tabular}[c]{@{}c@{}}$5\times 10^5$spins\\ $dt=0.005ms$\end{tabular}                      & 0.18     & 6.11     \\ \hline
		\begin{tabular}[c]{@{}c@{}}$h=0.1\mu m$\\ $rtol=10^{-4}$, $atol=10^{-6}$\end{tabular}                  & 0.09                                            & 0.37                                        & \begin{tabular}[c]{@{}c@{}}$1\times 10^6$spins\\ $dt=0.01ms$\end{tabular}                      & 0.20     & 4.32     \\ \hline
		\begin{tabular}[c]{@{}c@{}}$h=0.1\mu m$\\ $rtol=10^{-5}$, $atol=10^{-7}$\end{tabular}                  & 0.05                                            & 0.11                                        & \begin{tabular}[c]{@{}c@{}}$1\times 10^6$spins\\ $dt=0.005ms$\end{tabular}                      & 0.47     & 4.17     \\ \hline
		\begin{tabular}[c]{@{}c@{}}$h=0.05\mu m$\\ $rtol=10^{-6}$, $atol=10^{-8}$\end{tabular}                  & 0.04                                            & 0.06                                        & \begin{tabular}[c]{@{}c@{}}$2\times 10^6$spins\\ $dt=0.005ms$\end{tabular}                      & 0.19     & 1.07     \\ \hline
		\begin{tabular}[c]{@{}c@{}}$h=0.01\mu m$\\ $rtol=10^{-7}$, $atol=10^{-9}$\end{tabular}                  & 0.02                                            & 0.03                                        & \begin{tabular}[c]{@{}c@{}}$2\times 10^6$spins\\ $dt=0.001ms$\end{tabular}                      & 0.67     & 0.95     \\ \hline
	\end{tabular}
	\caption{The computational time (in seconds) and the maximum relative error (in percent) of SpinDoctor and GPU Monte-Carlo simulations for a rectangular cuboid ($L_x = 3\mu m, L_y = 100\mu m, L_z = 1\mu m$).  The maximum relative error $E_{max}$ is taken over 10 gradient directions uniformly placed on the unit semi-circle in the $x-y$ plane. The reference signal is the analytical Matrix Formalism signal.\label{table:accuracy_rectangle}}
\end{table}

From the rectangle cuboid example, we see that refining the SpinDoctor simulation parameters clearly results in a reduction in the simulation error, therefore we took the SpinDoctor simulation with the finest space discretization ({\textbf{Space-3}}: $Htetgen = 0.05\mu m$) and the finest time discretization
({\textbf{Time-4}}: $rtol = 10^{-6}$, $atol = 10^{-8}$) as the reference solution for the following neuron simulations we performed.
We remind the reader that by looking at the difference between the reference solution and the coarser SpinDoctor simulations in Figure \ref{fig:SD_validation}, we estimate that the accuracy of the reference solution is within 0.05\% of the true solution 
at $b=1000\bunit$ and it is within 0.2\% of the true solution at $b=4000\bunit$, for PGSE ($\delta = 10\tunit$, $\Delta = 43\tunit$).

\begin{figure}[H]
	\centering
	\includegraphics[width=1\textwidth]{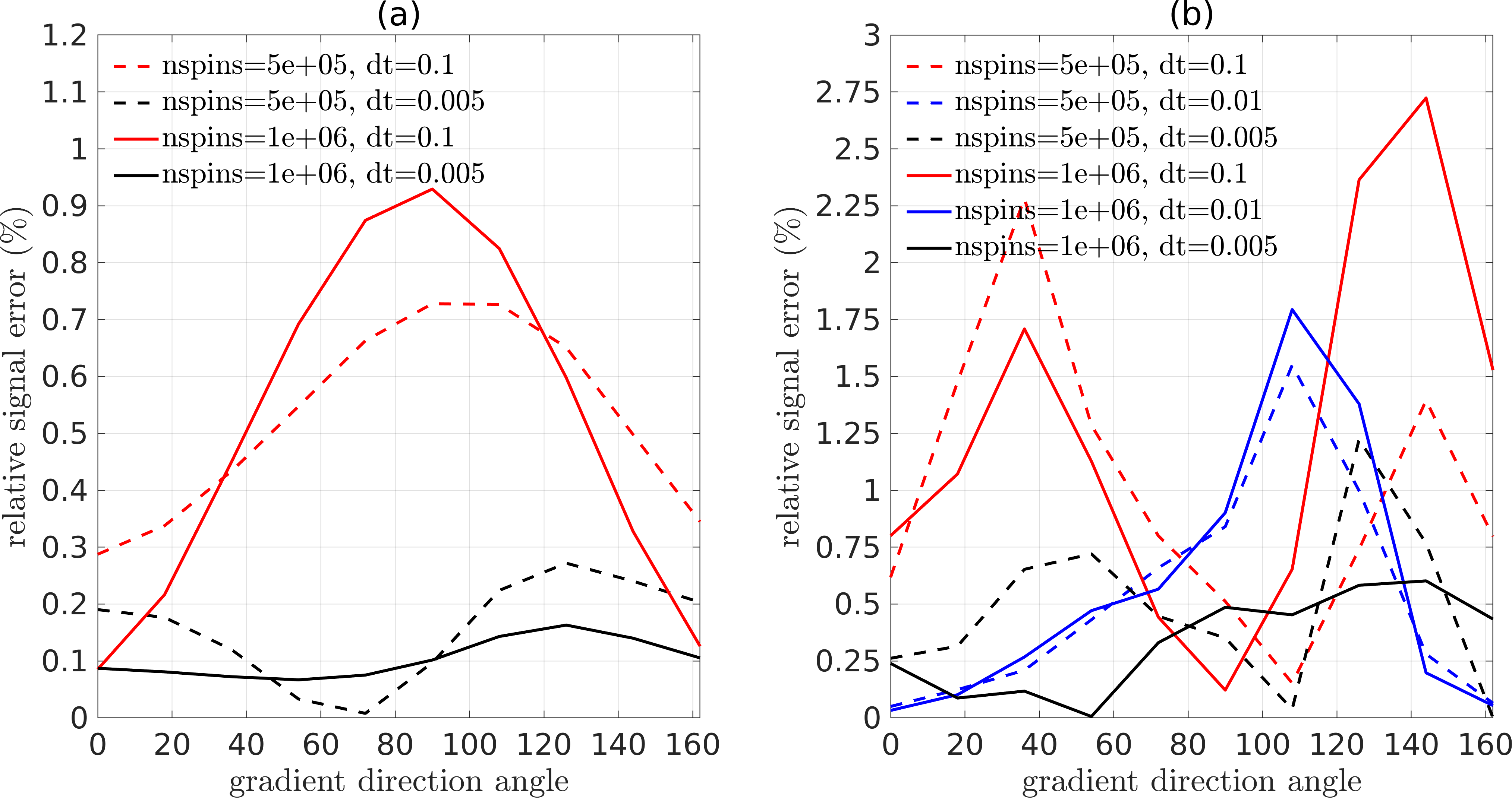}
	\caption{The relative signal difference between the SpinDoctor reference signal and the signals given by GPU Monte-Carlo simulations for the neuron {\it 03b\_spindle4aACC}.  The diffusion coefficient is $2 \times 10^{-3}\dunit$.  10 gradient directions uniformly placed on the
		unit semi-circle in the $x-y$ plane were simulated.  The gradient direction angle is given with respect to the $x$-axis.  
		The gradient sequence is PGSE $(\delta=10ms, \Delta=43ms)$.  (a) $b=1000\bunit$. (b) $b=4000\bunit$.
		\label{fig:GPU_validation}
	}
\end{figure}

In Figure \ref{fig:GPU_validation} we show the relative signal errors E (in percent) between several GPU MC simulations and the SpinDoctor reference solution.
Figure \ref{fig:GPU_validation}(a) shows that for $b=1000\bunit$, by refining the time discretization
from \textbf{MC Time-1} to \textbf{MC Time-3}, the $E_{max}$ went from around 0.7\%
down to 0.28\% for space discretization \textbf{MC Space-1} and from 0.92\% to 0.17\% for space discretization \textbf{MC Space-2}.  We see in Figure \ref{fig:GPU_validation}(b) that by using
\textbf{MC Space-2} and \textbf{MC Time-3},
% \be{}
% 5\times 10^{5} \text{ spins}, dt = 0.1\tunit;
% \label{eq:gpu_simulation_parameters}
% \ee
the $E_{max}$ is around $0.6\%$ for $b=4000\bunit$. 

\subsection{Timing}
\label{sec:simul_timing}

Here we compare the computational times used by
SpinDoctor and the GPU Monte-Carlo program \cite{Nguyen2018a} for neuron and cell components simulations.
The following are the computational platforms used for each type of simulation:
\begin{itemize}
	\item
	      All SpinDoctor simulations were performed on a Dell workstation (Intel(R) Xeon(R) CPU E5-2667 0 @ 2.90GHz, 189GB DDR4 RAM), running CentOS 7.4.1708;
	\item
	      All GPU Monte-Carlo simulations were performed on a Dell workstation (Intel(R) Xeon(R) CPU E5-2698 v4 @ 2.20GHz, 256GB DDR4 RAM and Tesla V100 DGXS 32GB), running Ubuntu 18.04.3 LTS.
\end{itemize}

For a fair comparison, we will choose two set of simulation parameters
with a comparable accuracy, in other words, a comparable $E_{max}$, 
this choice will be different for lower b-values versus higher b-values.   
Because the GPU Monte-Carlo simulation is inherently parallel (there is no communication between spins), the running time is essentially the same whether one runs one gradient direction or multiple directions, up to the limit of GPU memory.  We found that the GPU memory limit in the computer listed above to be about $1\times10^6$ spins and 1200 gradient directions running in parallel.  A rough estimation of the GPU limitation for Tesla V100 with 32GB GPU memory is that the number of spins times the number of gradient direction should be less than $1\times10^9$. In contrast, SpinDoctor is mostly designed for serial computations, so simulations for multiple gradient directions are run one after another.
Due to this difference between serial and parallel implementations, we show in Table \ref{tab:computational_time}
the SpinDoctor and GPU Monte-Carlo computational times for one gradient direction.  To estimate the SpinDoctor computational times one can multiply the computational time in  Table \ref{tab:computational_time} by the number of gradient directions to be simulated.

From Table \ref{tab:computational_time}, we see that the ratio between the GPU Monte-Carlo computational time and the SpinDoctor computational time for 1 gradient 
direction ranges from 31 to 72 for the whole neuron simulations.  This means that 31 (for $b=1000\bunit$) and 72 (for $b=4000\bunit$) 
gradient directions are the break-even point when considering whether to run GPU Monte-Carlo or the SpinDoctor code in terms of 
computational time.  The break-even point is between 24 to 43 gradient directions for the dendrite branches, and between 46 to 55 gradient 
directions for the soma.  Details of the accuracy and the computational time of 6 SpinDoctor and 6 GPU Monte-Carlo simulations can be 
found in Tables \ref{tab:time_error_neuron} - \ref{tab:time_error_soma}.

\begin{table}[H]
	\centering
	\begin{tabular}{|c|c|c|c|c|c|c|c|c|c|}
		\hline
		\multicolumn{2}{|c|}{Computational time (s)} & \multicolumn{2}{c|}{neuron} & \multicolumn{2}{c|}{dendrite1} & \multicolumn{2}{c|}{dendrite2} & \multicolumn{2}{c|}{soma}                                                                          \\ \hline
		$\delta=10ms, \Delta=43ms$                   & $b$                         & $E_{max}$                      & $t\ (s)$                       & $E_{max}$                 & $t\ (s)$ & $E_{max}$         & $t\ (s)$ & $E_{max}$         & $t\ (s)$ \\ \hline
		\multirow{2}{*}{SpinDoctor}                  & $1000$                      & 0.16                           & 17.1                           & 0.52                      & 7.8      & 0.11              & 3.4      & 0.17              & 5.7      \\ \cline{2-10}
		                                             & $4000$                      & 0.34                           & 26.0                           & 0.91                      & 11.8     & 0.13              & 4.9      & 0.23              & 6.8      \\ \hline
		\multirow{2}{*}{GPU Monte-Carlo}             & $1000$                      & 0.14                           & 537.8                          & 0.65                      & 337.3    & 0.51              & 116.1    & 0.29              & 311.2    \\ \cline{2-10}
		                                             & $4000$                      & 0.60                           & 1895.9                         & 2.34                      & 342.1    & 0.47              & 116.5    & 0.90              & 311.3    \\ \hline
		\multirow{2}{*}{\begin{tabular}[c]{@{}c@{}}GPU MC/\\ SpinDoctor ratio\end{tabular}}  & $1000$                      & \multirow{2}{*}{}              & 31                             & \multirow{2}{*}{}         & 43       & \multirow{2}{*}{} & 34       & \multirow{2}{*}{} & 55       \\ \cline{2-2} \cline{4-4} \cline{6-6} \cline{8-8} \cline{10-10}
		                                             & $4000$                      &                                & 72                             &                           & 29*      &                   & 24       &                   & 46       \\ \hline
	\end{tabular}
	\caption{The computational times of SpinDoctor and GPU Monte-Carlo simulations (in seconds). The simulation parameters for SpinDoctor are $rtol = 10^{-3}, atol = 10^{-5}, Htetgen = 0.5\mu m$. The space discretization is: dendrite1 (finite elements mesh: 10825 nodes and 30617 elements), dendrite2 (finite elements mesh: 6444 nodes and 24051 elements), the soma (finite elements mesh: 5842 nodes and 26160 elements), the whole neuron (finite elements mesh: 19425 nodes and 60431 elements). The simulation parameters for GPU Monte-Carlo can be found in Tables \ref{tab:time_error_neuron} - \ref{tab:time_error_soma}. The maximum relative error $E_{max}$ (in percent) is taken over 10 gradient directions
		uniformly placed on the unit semi-circle in the $x-y$ plane.  *: The GPU MC error 2.34\% is too large, this computational time ratio should not be used.}
	\label{tab:computational_time}
\end{table}

\begin{table}[H]
	\centering
	\begin{tabular}{|c|c|c|c|c|c|c|c|c|c|}
		\hline
		\multirow{3}{*}{\begin{tabular}[c]{@{}c@{}}SpinDoctor\\ $\delta=10ms$\\ $\Delta=43ms$\end{tabular}} & \multicolumn{2}{c|}{\multirow{2}{*}{$b=1000$}} & \multicolumn{2}{c|}{\multirow{2}{*}{$b=4000$}} & \multirow{3}{*}{\begin{tabular}[c]{@{}c@{}}GPU MC\\ $\delta=10ms$\\ $\Delta=43ms$\end{tabular}} & \multicolumn{2}{c|}{\multirow{2}{*}{$b=1000$}} & \multicolumn{2}{c|}{\multirow{2}{*}{$b=4000$}}                                                                                                                \\
		                                            & \multicolumn{2}{c|}{}                          & \multicolumn{2}{c|}{}                          &                                             & \multicolumn{2}{c|}{}                          & \multicolumn{2}{c|}{}                                                                                                                                         \\ \cline{2-5} \cline{7-10}
		                                            & $E_{max}$                                      & $t\ (s)$                                       & $E_{max}$                                   & $t\ (s)$                                       &                                                & $E_{max}$                           & $t\ (s)$       & $E_{max}$                           & $t\ (s)$        \\ \hline
		\begin{tabular}[c]{@{}c@{}}$\mathbf{h=0.5\mu m}$\\ $\mathbf{rtol=10^{-3}}$\\ $\mathbf{atol=10^{-5}}$\end{tabular}                  & \textbf{0.16}                                  & \textbf{17.1}                                  & \textbf{0.34}                               & \textbf{26.0}                                  & \begin{tabular}[c]{@{}c@{}}$5\times 10^5spins$\\ $dt=0.1ms$\end{tabular}                     & \begin{tabular}[c]{@{}c@{}}0.73\\ (0.87)\end{tabular}          & 89.1           & \begin{tabular}[c]{@{}c@{}}2.28\\ (2.40)\end{tabular}          & 89.7            \\ \hline
		\begin{tabular}[c]{@{}c@{}}$h=0.5\mu m$\\ $rtol=10^{-4}$\\ $atol=10^{-6}$\end{tabular}                  & 0.18                                           & 26.6                                           & 0.59                                        & 39.6                                           & \begin{tabular}[c]{@{}c@{}}$\mathbf{5\times 10^5spins}$\\ $\mathbf{dt=0.01ms}$\end{tabular}                     & \textbf{\begin{tabular}[c]{@{}c@{}}0.14\\ (0.08)\end{tabular}} & \textbf{537.8} & \begin{tabular}[c]{@{}c@{}}1.55\\ (2.01)\end{tabular}          & 538.4           \\ \hline
		\begin{tabular}[c]{@{}c@{}}$h=0.1\mu m$\\ $rtol=10^{-3}$\\ $atol=10^{-5}$\end{tabular}                  & 0.06                                           & 93.5                                           & 0.10                                        & 137.8                                          & \begin{tabular}[c]{@{}c@{}}$5\times 10^5spins$\\ $dt=0.005ms$\end{tabular}                     & \begin{tabular}[c]{@{}c@{}}0.27\\ (0.11)\end{tabular}          & 961.9          & \begin{tabular}[c]{@{}c@{}}1.22\\ (0.84)\end{tabular}          & 966.9           \\ \hline
		\begin{tabular}[c]{@{}c@{}}$h=0.1\mu m$\\ $rtol=10^{-4}$\\ $atol=10^{-6}$\end{tabular}                  & 0.05                                           & 190.4                                          & 0.20                                        & 274.9                                          & \begin{tabular}[c]{@{}c@{}}$1\times 10^6spins$\\ $dt=0.1ms$\end{tabular}                     & \begin{tabular}[c]{@{}c@{}}0.93\\ (1.03)\end{tabular}          & 172.5          & \begin{tabular}[c]{@{}c@{}}2.72\\ (2.13)\end{tabular}          & 172.8           \\ \hline
		\begin{tabular}[c]{@{}c@{}}$h=0.1\mu m$\\ $rtol=10^{-5}$\\ $atol=10^{-7}$\end{tabular}                  & 0.05                                           & 158.8                                          & 0.15                                        & 228.0                                          & \begin{tabular}[c]{@{}c@{}}$1\times 10^6spins$\\ $dt=0.01ms$\end{tabular}                     & \begin{tabular}[c]{@{}c@{}}0.11\\ (0.09)\end{tabular}          & 1052.6         & \begin{tabular}[c]{@{}c@{}}1.79\\ (2.26)\end{tabular}          & 1056.9          \\ \hline
		\begin{tabular}[c]{@{}c@{}}$h=0.05\mu m$\\ $rtol=10^{-6}$\\ $atol=10^{-8}$\end{tabular}                  & ref.                                           & 523.0                                          & ref.                                        & 806.6                                          & \begin{tabular}[c]{@{}c@{}}$\mathbf{1\times 10^6spins}$\\ $\mathbf{dt=0.005ms}$\end{tabular}                     & \begin{tabular}[c]{@{}c@{}}0.16\\ (ref.)\end{tabular}          & 1896.6         & \textbf{\begin{tabular}[c]{@{}c@{}}0.60\\ (ref.)\end{tabular}} & \textbf{1895.9} \\ \hline
	\end{tabular}
	\caption{The computational time (in seconds) and the maximum relative error (in percent) of SpinDoctor and GPU Monte-Carlo simulations for the neuron {\it 03b\_spindle4aACC}. The maximum relative error $E_{max}$ is taken over 10 gradient directions uniformly placed on the unit semi-circle in the $x-y$ plane. For the $E_{max}$ of SpinDoctor, the reference signal is the one with the finest space discretization and the smallest time discretization, i.e. $h=0.05 \mu m, rtol=10^{-6}, atol=10^{-8}$. For the $E_{max}$ of GPU Monte-Carlo, two reference signals are used, one is the signal given by SpinDoctor with $h=0.05 \mu m, rtol=10^{-6}, atol=10^{-8}$, the other is the signal given by GPU Monte-Carlo with $10^{6}\ spins$ and $dt=0.005ms$ ($E_{max}$ for this case is written in the parenthesis). The data in bold are used in Table \ref{tab:computational_time}.}
	\label{tab:time_error_neuron}
\end{table}

\begin{table}[H]
	\centering
	\begin{tabular}{|c|c|c|c|c|c|c|c|c|c|}
		\hline
		\multirow{3}{*}{\begin{tabular}[c]{@{}c@{}}SpinDoctor\\ $\delta=10ms$\\ $\Delta=43ms$\end{tabular}} & \multicolumn{2}{c|}{\multirow{2}{*}{$b=1000$}} & \multicolumn{2}{c|}{\multirow{2}{*}{$b=4000$}} & \multirow{3}{*}{\begin{tabular}[c]{@{}c@{}}GPU MC\\ $\delta=10ms$\\ $\Delta=43ms$\end{tabular}} & \multicolumn{2}{c|}{\multirow{2}{*}{$b=1000$}} & \multicolumn{2}{c|}{\multirow{2}{*}{$b=4000$}}                                                                                                               \\
		                                            & \multicolumn{2}{c|}{}                          & \multicolumn{2}{c|}{}                          &                                             & \multicolumn{2}{c|}{}                          & \multicolumn{2}{c|}{}                                                                                                                                        \\ \cline{2-5} \cline{7-10}
		                                            & $E_{max}$                                      & $t\ (s)$                                       & $E_{max}$                                   & $t\ (s)$                                       &                                                & $E_{max}$                           & $t\ (s)$       & $E_{max}$                           & $t\ (s)$       \\ \hline
		\begin{tabular}[c]{@{}c@{}}$\mathbf{h=0.5\mu m}$\\ $\mathbf{rtol=10^{-3}}$\\ $\mathbf{atol=10^{-5}}$\end{tabular}                  & \textbf{0.52}                                  & \textbf{7.8}                                   & \textbf{0.91}                               & \textbf{11.8}                                  & \begin{tabular}[c]{@{}c@{}}$5\times 10^5spins$\\ $dt=0.1ms$\end{tabular}                     & \begin{tabular}[c]{@{}c@{}}1.36\\ (0.26)\end{tabular}          & 173.9          & \begin{tabular}[c]{@{}c@{}}1.35\\ (1.47)\end{tabular}          & 174.7          \\ \hline
		\begin{tabular}[c]{@{}c@{}}$h=0.5\mu m$\\ $rtol=10^{-4}$\\ $atol=10^{-6}$\end{tabular}                  & 0.65                                           & 7.8                                            & 1.03                                        & 11.4                                           & \begin{tabular}[c]{@{}c@{}}$5\times 10^5spins$\\ $dt=0.01ms$\end{tabular}                     & \begin{tabular}[c]{@{}c@{}}1.48\\ (0.18)\end{tabular}          & 715.3          & \begin{tabular}[c]{@{}c@{}}3.23\\ (1.03)\end{tabular}          & 715.2          \\ \hline
		\begin{tabular}[c]{@{}c@{}}$h=0.1\mu m$\\ $rtol=10^{-3}$\\ $atol=10^{-5}$\end{tabular}                  & 0.10                                           & 13.1                                           & 0.42                                        & 19.6                                           & \begin{tabular}[c]{@{}c@{}}$5\times 10^5spins$\\ $dt=0.005ms$\end{tabular}                     & \begin{tabular}[c]{@{}c@{}}1.44\\ (0.19)\end{tabular}          & 1487.3         & \begin{tabular}[c]{@{}c@{}}2.78\\ (1.86)\end{tabular}          & 1488.9         \\ \hline
		\begin{tabular}[c]{@{}c@{}}$h=0.1\mu m$\\ $rtol=10^{-4}$\\ $atol=10^{-6}$\end{tabular}                  & 0.22                                           & 21.4                                           & 0.34                                        & 32.2                                           & \begin{tabular}[c]{@{}c@{}}$\mathbf{1\times 10^6spins}$\\ $\mathbf{dt=0.1ms}$\end{tabular}                     & \textbf{\begin{tabular}[c]{@{}c@{}}0.65\\ (0.98)\end{tabular}} & \textbf{337.3} & \textbf{\begin{tabular}[c]{@{}c@{}}2.34\\ (0.46)\end{tabular}} & \textbf{342.1} \\ \hline
		\begin{tabular}[c]{@{}c@{}}$h=0.1\mu m$\\ $rtol=10^{-5}$\\ $atol=10^{-7}$\end{tabular}                  & 0.17                                           & 22.6                                           & 0.24                                        & 31.7                                           & \begin{tabular}[c]{@{}c@{}}$1\times 10^6spins$\\ $dt=0.01ms$\end{tabular}                     & \begin{tabular}[c]{@{}c@{}}1.29\\ (0.33)\end{tabular}          & 1401.3         & \begin{tabular}[c]{@{}c@{}}1.24\\ (1.55)\end{tabular}          & 1401.1         \\ \hline
		\begin{tabular}[c]{@{}c@{}}$h=0.05\mu m$\\ $rtol=10^{-6}$\\ $atol=10^{-8}$\end{tabular}                  & ref.                                           & 65.1                                           & ref.                                        & 103.3                                          & \begin{tabular}[c]{@{}c@{}}$1\times 10^6spins$\\ $dt=0.005ms$\end{tabular}                     & \begin{tabular}[c]{@{}c@{}}1.63\\ (ref.)\end{tabular}          & 2952.9         & \begin{tabular}[c]{@{}c@{}}2.23\\ (ref.)\end{tabular}          & 2923.7         \\ \hline
	\end{tabular}
	\caption{The computational time (in seconds) and the maximum relative error (in percent) of SpinDoctor and GPU Monte-Carlo simulations for the dendrite {\it 03b\_spindle4aACC\_dendrites\_1}. The maximum relative error $E_{max}$ is taken over 10 gradient directions uniformly placed on the unit semi-circle in the $x-y$ plane. For the $E_{max}$ of SpinDoctor, the reference signal is the one with the finest space discretization and the smallest time discretization, i.e. $h=0.05 \mu m, rtol=10^{-6}, atol=10^{-8}$. For the $E_{max}$ of GPU Monte-Carlo, two reference signals are used, one is the signal given by SpinDoctor with $h=0.05 \mu m, rtol=10^{-6}, atol=10^{-8}$, the other is the signal given by GPU Monte-Carlo with $10^{6}\ spins$ and $dt=0.005ms$ ($E_{max}$ for this case is written in the parenthesis). The data in bold are used in Table \ref{tab:computational_time}.}
	\label{tab:time_error_dendrite1}
\end{table}

\begin{table}[H]
	\centering
	\begin{tabular}{|c|c|c|c|c|c|c|c|c|c|}
		\hline
		\multirow{3}{*}{\begin{tabular}[c]{@{}c@{}}SpinDoctor\\ $\delta=10ms$\\ $\Delta=43ms$\end{tabular}} & \multicolumn{2}{c|}{\multirow{2}{*}{$b=1000$}} & \multicolumn{2}{c|}{\multirow{2}{*}{$b=4000$}} & \multirow{3}{*}{\begin{tabular}[c]{@{}c@{}}GPU MC\\ $\delta=10ms$\\ $\Delta=43ms$\end{tabular}} & \multicolumn{2}{c|}{\multirow{2}{*}{$b=1000$}} & \multicolumn{2}{c|}{\multirow{2}{*}{$b=4000$}}                                                                                                                \\
		                                            & \multicolumn{2}{c|}{}                          & \multicolumn{2}{c|}{}                          &                                             & \multicolumn{2}{c|}{}                          & \multicolumn{2}{c|}{}                                                                                                                                         \\ \cline{2-5} \cline{7-10}
		                                            & $E_{max}$                                      & $t\ (s)$                                       & $E_{max}$                                   & $t\ (s)$                                       &                                                & $E_{max}$                           & $t\ (s)$       & $E_{max}$                            & $t\ (s)$       \\ \hline
		\begin{tabular}[c]{@{}c@{}}$\mathbf{h=0.5\mu m}$\\ $\mathbf{rtol=10^{-3}}$\\ $\mathbf{atol=10^{-5}}$\end{tabular}                  & \textbf{0.11}                                  & \textbf{3.4}                                   & \textbf{0.13}                               & \textbf{4.9}                                   & \begin{tabular}[c]{@{}c@{}}$\mathbf{5\times 10^5spins}$\\ $\mathbf{dt=0.1ms}$\end{tabular}                     & \textbf{\begin{tabular}[c]{@{}c@{}}0.51\\ (0.21)\end{tabular}} & \textbf{116.1} & \textbf{\begin{tabular}[c]{@{}c@{}}0.47\\ (3.19)\end{tabular}} & \textbf{116.5} \\ \hline
		\begin{tabular}[c]{@{}c@{}}$h=0.5\mu m$\\ $rtol=10^{-4}$\\ $atol=10^{-6}$\end{tabular}                 & 0.11                                           & 5.8                                            & 0.24                                        & 8.1                                            & \begin{tabular}[c]{@{}c@{}}$5\times 10^5spins$\\ $dt=0.01ms$\end{tabular}                    & \begin{tabular}[c]{@{}c@{}}0.83\\ (0.13)\end{tabular}         & 638.5          & \begin{tabular}[c]{@{}c@{}}1.01\\ (4.36)\end{tabular}          & 638.8          \\ \hline
		\begin{tabular}[c]{@{}c@{}}$h=0.1\mu m$\\ $rtol=10^{-3}$\\ $atol=10^{-5}$\end{tabular}                 & 0.08                                           & 6.5                                            & 0.19                                        & 8.9                                            & \begin{tabular}[c]{@{}c@{}}$5\times 10^5spins$\\ $dt=0.005ms$\end{tabular}                    & \begin{tabular}[c]{@{}c@{}}0.69\\ (0.04)\end{tabular}         & 1174.8         & \begin{tabular}[c]{@{}c@{}}2.40\\ (1.69)\end{tabular}          & 1174.7         \\ \hline
		\begin{tabular}[c]{@{}c@{}}$h=0.1\mu m$\\ $rtol=10^{-4}$\\ $atol=10^{-6}$\end{tabular}                 & 0.05                                           & 9.0                                            & 0.07                                        & 12.7                                           & \begin{tabular}[c]{@{}c@{}}$1\times 10^6spins$\\ $dt=0.1ms$\end{tabular}                    & \begin{tabular}[c]{@{}c@{}}0.85\\ (0.14)\end{tabular}         & 226.1          & \begin{tabular}[c]{@{}c@{}}1.30\\ (2.48)\end{tabular}          & 226.6          \\ \hline
		\begin{tabular}[c]{@{}c@{}}$h=0.1\mu m$\\ $rtol=10^{-5}$\\ $atol=10^{-7}$\end{tabular}                 & 0.07                                           & 12.1                                           & 0.11                                        & 16.8                                           & \begin{tabular}[c]{@{}c@{}}$1\times 10^6spins$\\ $dt=0.01ms$\end{tabular}                    & \begin{tabular}[c]{@{}c@{}}0.74\\ (0.04)\end{tabular}         & 1260.1         & \begin{tabular}[c]{@{}c@{}}0.91\\ (2.83)\end{tabular}          & 1265.7         \\ \hline
		\begin{tabular}[c]{@{}c@{}}$h=0.05\mu m$\\ $rtol=10^{-6}$\\ $atol=10^{-8}$\end{tabular}                 & ref.                                           & 22.0                                           & ref.                                        & 31.8                                           & \begin{tabular}[c]{@{}c@{}}$1\times 10^6spins$\\ $dt=0.005ms$\end{tabular}                    & \begin{tabular}[c]{@{}c@{}}0.72\\ (ref.)\end{tabular}         & 2339.1         & \begin{tabular}[c]{@{}c@{}}3.56\\ (ref.)\end{tabular}          & 2320.7         \\ \hline
	\end{tabular}
	\caption{The computational time (in seconds) and the maximum relative error (in percent) of SpinDoctor and GPU Monte-Carlo simulations for the dendrite {\it 03b\_spindle4aACC\_dendrites\_2}. The maximum relative error $E_{max}$ is taken over 10 gradient directions uniformly placed on the unit semi-circle in the $x-y$ plane. For the $E_{max}$ of SpinDoctor, the reference signal is the one with the finest space discretization and the smallest time discretization, i.e. $h=0.05 \mu m, rtol=10^{-6}, atol=10^{-8}$. For the $E_{max}$ of GPU Monte-Carlo, two reference signals are used, one is the signal given by SpinDoctor with $h=0.05 \mu m, rtol=10^{-6}, atol=10^{-8}$, the other is the signal given by GPU Monte-Carlo with $10^{6}\ spins$ and $dt=0.005ms$ ($E_{max}$ for this case is written in the parenthesis). The data in bold are used in Table \ref{tab:computational_time}.}
	\label{tab:time_error_dendrite2}
\end{table}

\begin{table}[H]
	\centering
	\begin{tabular}{|c|c|c|c|c|c|c|c|c|c|}
		\hline
		\multirow{3}{*}{\begin{tabular}[c]{@{}c@{}}SpinDoctor\\ $\delta=10ms$\\ $\Delta=43ms$\end{tabular}} & \multicolumn{2}{c|}{\multirow{2}{*}{$b=1000$}} & \multicolumn{2}{c|}{\multirow{2}{*}{$b=4000$}} & \multirow{3}{*}{\begin{tabular}[c]{@{}c@{}}GPU MC\\ $\delta=10ms$\\ $\Delta=43ms$\end{tabular}} & \multicolumn{2}{c|}{\multirow{2}{*}{$b=1000$}} & \multicolumn{2}{c|}{\multirow{2}{*}{$b=4000$}}                                                                                                                 \\
		                                             & \multicolumn{2}{c|}{}                          & \multicolumn{2}{c|}{}                          &                                              & \multicolumn{2}{c|}{}                          & \multicolumn{2}{c|}{}                                                                                                                                          \\ \cline{2-5} \cline{7-10}
		                                             & $E_{max}$                                      & $t\ (s)$                                       & $E_{max}$                                    & $t\ (s)$                                       &                                                & $E_{max}$                            & $t\ (s)$       & $E_{max}$                            & $t\ (s)$       \\ \hline
		\begin{tabular}[c]{@{}c@{}}$\mathbf{h=0.5\mu m}$\\ $\mathbf{rtol=10^{-3}}$\\ $\mathbf{atol=10^{-5}}$\end{tabular}                  & \textbf{0.17}                                  & \textbf{5.7}                                   & \textbf{0.23}                                & \textbf{6.8}                                   & \begin{tabular}[c]{@{}c@{}}$5\times 10^5spins$\\ $dt=0.1ms$\end{tabular}                    & \begin{tabular}[c]{@{}c@{}}1.46\\ (1.31)\end{tabular}          & 52.8           & \begin{tabular}[c]{@{}c@{}}4.16\\ (4.75)\end{tabular}          & 52.6           \\ \hline
		\begin{tabular}[c]{@{}c@{}}$h=0.5\mu m$\\ $rtol=10^{-4}$\\ $atol=10^{-6}$\end{tabular}                  & 0.06                                           & 8.6                                            & 0.23                                         & 10.3                                           & \begin{tabular}[c]{@{}c@{}}$\mathbf{5\times 10^5spins}$\\ $\mathbf{dt=0.01ms}$\end{tabular}                    & \textbf{\begin{tabular}[c]{@{}c@{}}0.29\\ (0.20)\end{tabular}} & \textbf{311.2} & \textbf{\begin{tabular}[c]{@{}c@{}}0.90\\ (1.50)\end{tabular}} & \textbf{311.3} \\ \hline
		\begin{tabular}[c]{@{}c@{}}$h=0.1\mu m$\\ $rtol=10^{-3}$\\ $atol=10^{-5}$\end{tabular}                  & 0.24                                           & 29.7                                           & 0.55                                         & 34.2                                           & \begin{tabular}[c]{@{}c@{}}$5\times 10^5spins$\\ $dt=0.005ms$\end{tabular}                    & \begin{tabular}[c]{@{}c@{}}0.09\\ (0.23)\end{tabular}          & 564.9          & \begin{tabular}[c]{@{}c@{}}2.34\\ (3.23)\end{tabular}          & 565.2          \\ \hline
		\begin{tabular}[c]{@{}c@{}}$h=0.1\mu m$\\ $rtol=10^{-4}$\\ $atol=10^{-6}$\end{tabular}                  & 0.04                                           & 36.7                                           & 0.11                                         & 43.7                                           & \begin{tabular}[c]{@{}c@{}}$1\times 10^6spins$\\ $dt=0.1ms$\end{tabular}                    & \begin{tabular}[c]{@{}c@{}}1.12\\ (0.99)\end{tabular}          & 101.9          & \begin{tabular}[c]{@{}c@{}}3.75\\ (3.72)\end{tabular}          & 102.3          \\ \hline
		\begin{tabular}[c]{@{}c@{}}$h=0.1\mu m$\\ $rtol=10^{-5}$\\ $atol=10^{-7}$\end{tabular}                  & 0.02                                           & 55.8                                           & 0.12                                         & 64.5                                           & \begin{tabular}[c]{@{}c@{}}$1\times 10^6spins$\\ $dt=0.01ms$\end{tabular}                    & \begin{tabular}[c]{@{}c@{}}0.1\\ (0.07)\end{tabular}          & 612.9          & \begin{tabular}[c]{@{}c@{}}1.37\\ (1.61)\end{tabular}          & 611.4          \\ \hline
		\begin{tabular}[c]{@{}c@{}}$h=0.05\mu m$\\ $rtol=10^{-6}$\\ $atol=10^{-8}$\end{tabular}                  & ref.                                           & 222                                            & ref.                                         & 247.9                                          & \begin{tabular}[c]{@{}c@{}}$1\times 10^6spins$\\ $dt=0.005ms$\end{tabular}                    & \begin{tabular}[c]{@{}c@{}}0.16\\ (ref.)\end{tabular}          & 1112.5         & \begin{tabular}[c]{@{}c@{}}1.06\\ (ref.)\end{tabular}          & 1110.1         \\ \hline
	\end{tabular}
	\caption{The computational time (in seconds) and the maximum relative error (in percent) of SpinDoctor and GPU Monte-Carlo simulations for the soma {\it 03b\_spindle4aACC\_soma}. The maximum relative error $E_{max}$ is taken over 10 gradient directions uniformly placed on the unit semi-circle in the $x-y$ plane. For the $E_{max}$ of SpinDoctor, the reference signal is the one with the finest space discretization and the smallest time discretization, i.e. $h=0.05 \mu m, rtol=10^{-6}, atol=10^{-8}$. For the $E_{max}$ of GPU Monte-Carlo, two reference signals are used, one is the signal given by SpinDoctor with $h=0.05 \mu m, rtol=10^{-6}, atol=10^{-8}$, the other is the signal given by GPU Monte-Carlo with $10^{6}\ spins$ and $dt=0.005ms$ ($E_{max}$ for this case is written in the parenthesis). The data in bold are used in Table \ref{tab:computational_time}.}
	\label{tab:time_error_soma}
\end{table}

\subsection{Choice of space and time discretization parameters for diffusion MRI simulations}

\label{sec:simul_parameters}
\begin{figure}[H]
	\centering
	\includegraphics[width=0.45\textwidth]{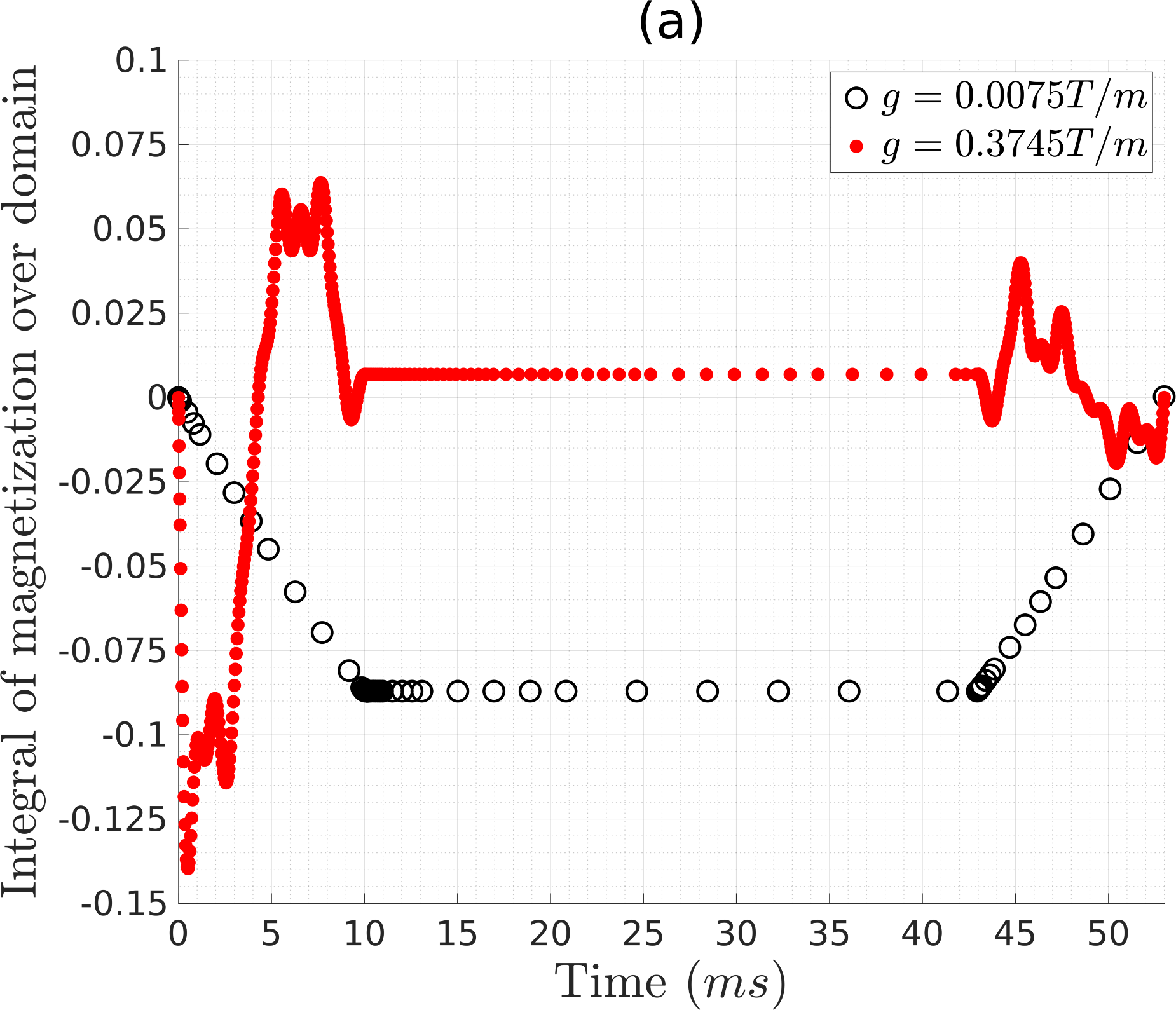}
	\includegraphics[width=0.45\textwidth]{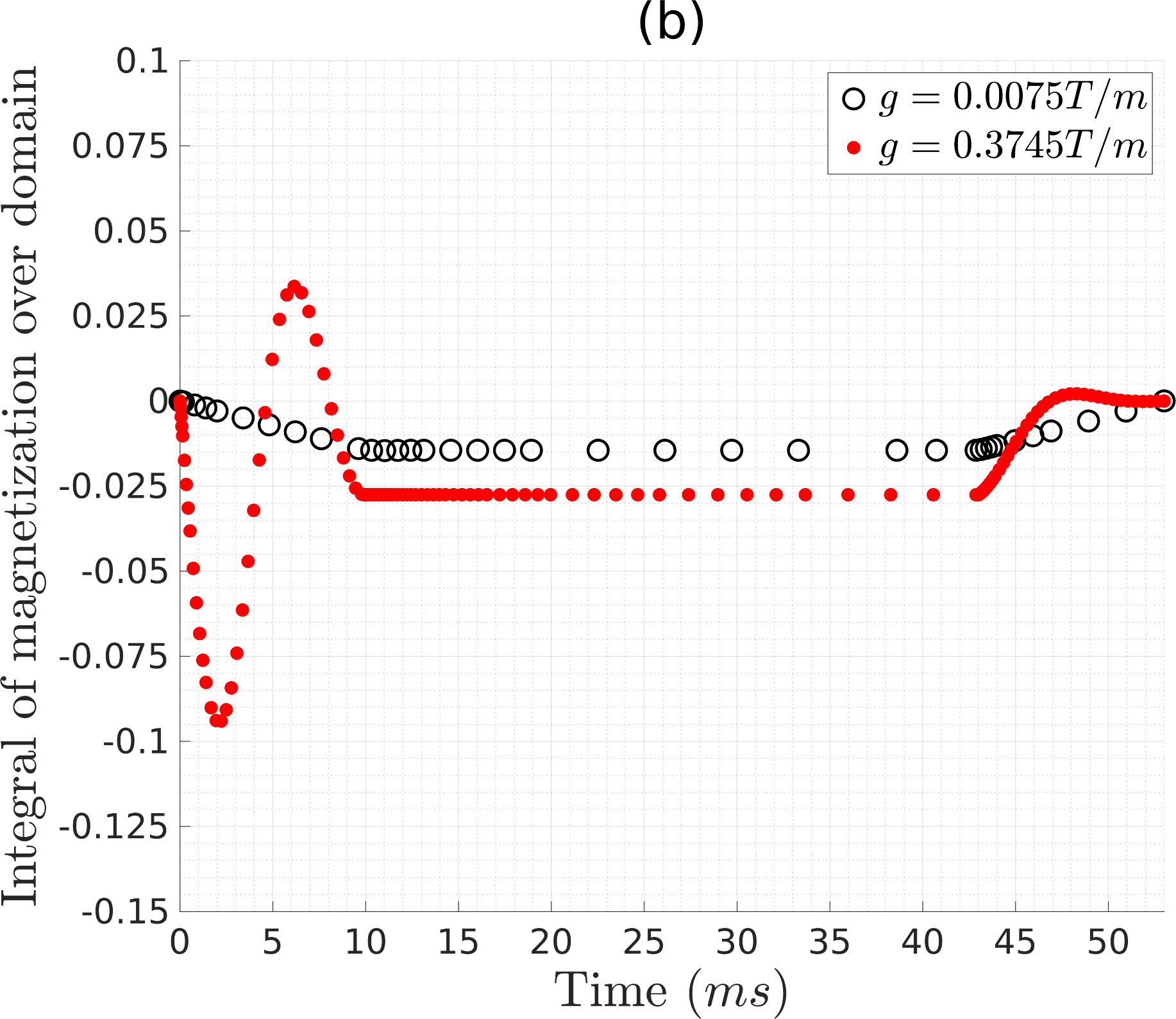}
	\caption{The integral of the imaginary part of the magnetization over the computational domain as a function
	of time.  The time discretization points chosen by SpinDoctor are indicated by the (time) positions of the markers.
	The experimental parameters are: PGSE ($\delta=10\tunit,\Delta=43\tunit$), gradient direction
	$\bug =[-0.3536   -0.6124   -0.7071]$.
	Two gradient amplitudes, $\vert \bg \vert = 0.0075 \qunit$  and $\vert \bg \vert = 0.3745 \qunit$ were simulated, equivalent to
	$b = 15.9\bunit$ and $b = 39666.7\bunit$, respectively.
	For these two b-values, SpinDoctor used a total of 54 (shown in black, left) and 506 (shown in red, left) non-uniformly spaced time steps to simulate the whole neuron;
	it took 38 (shown in black, right) and 145 (shown in red, right) non-uniformly spaced time steps to simulate its soma.
	(a) the neuron {\it 03b\_spindle4aACC}.  (b) its soma.}
	\label{fig:magnetization_oscillation}
\end{figure}

This section concerns the choice of discretization parameters for diffusion MRI simulations.
First we discuss time discretization. The Monte-Carlo implementations usually decide on
a time step $dt$ that is used throughout the entire simulation.  SpinDoctor uses residual tolerances
for the Matlab ODE solver to control the time step size.  The advantage of using residual
tolerances is that the time step is automatically made smaller when the magnetization
is oscillatory, and it is automatically made larger when the magnetization is smooth.  In
Figure \ref{fig:magnetization_oscillation} we show the SpinDoctor simulated magnetization solution
during the time interval $t\in[0,\delta+\Delta]$ for the PGSE sequence ($\delta=10\tunit$,
$\Delta=43\tunit$) for the full neuron and the soma, where we integrated the magnetization
over the computational domain, in this case, the whole neuron and the soma, respectively.  This integral
was evaluated at SpinDoctor time discretization points.  Because the magnetization is a complex-valued
quantity and the imaginary part of the magnetization \soutnew{}{, which encodes the spin phase information, }
is usually more oscillatory than the real part,
we show only the integral of the imaginary part of the magnetization.
\soutnew{}{We note during $(0,TE)$, the magnetization is complex-valued and both the real and imaginary parts are significant and contribute to the time-evolution.  After reforcusing, the imaginary part should be zero theoretically.  At $t = TE$, the non-zero imaginary
	part of the simulated magnetization is due to ``numerical discretization error'', whose size is related to the FE mesh size and the ODE solver tolerances.}

It is clear from Figure \ref{fig:magnetization_oscillation} that at the higher b-value the total magnetization
has more oscillations in time for both the whole neuron and the soma.
In addition, there are many more oscillations in time for the whole neuron than for the soma.
This justifies the SpinDoctor choice of using 54 (low gradient amplitude) and 506 (high gradient amplitude) non-uniformly spaced time steps to simulate the whole neuron, whereas it used
38 (low gradient amplitude) and 145 (high gradient amplitude) 
non-uniformly spaced time steps to simulate its soma.

To give an indication why more time discretization points are needed at higher gradient amplitudes,
we computed the eigenfunctions and eigenvalues of Bloch-Torrey operator, which governs the dynamics
of the magnetization during the gradient pulses ($t\in[0,\delta]$, $t\in[\Delta,\Delta+\delta]$, $\bg \neq 0$).
The Bloch-Torrey eigenfunctions and eigenvalues 
on the neuron and the soma are numerically computed using the Matrix
Formalism Module \cite{li2019practical} within the SpinDoctor Toolbox. 

\begin{figure}[H]
	\centering
	\includegraphics[width=0.49\textwidth]{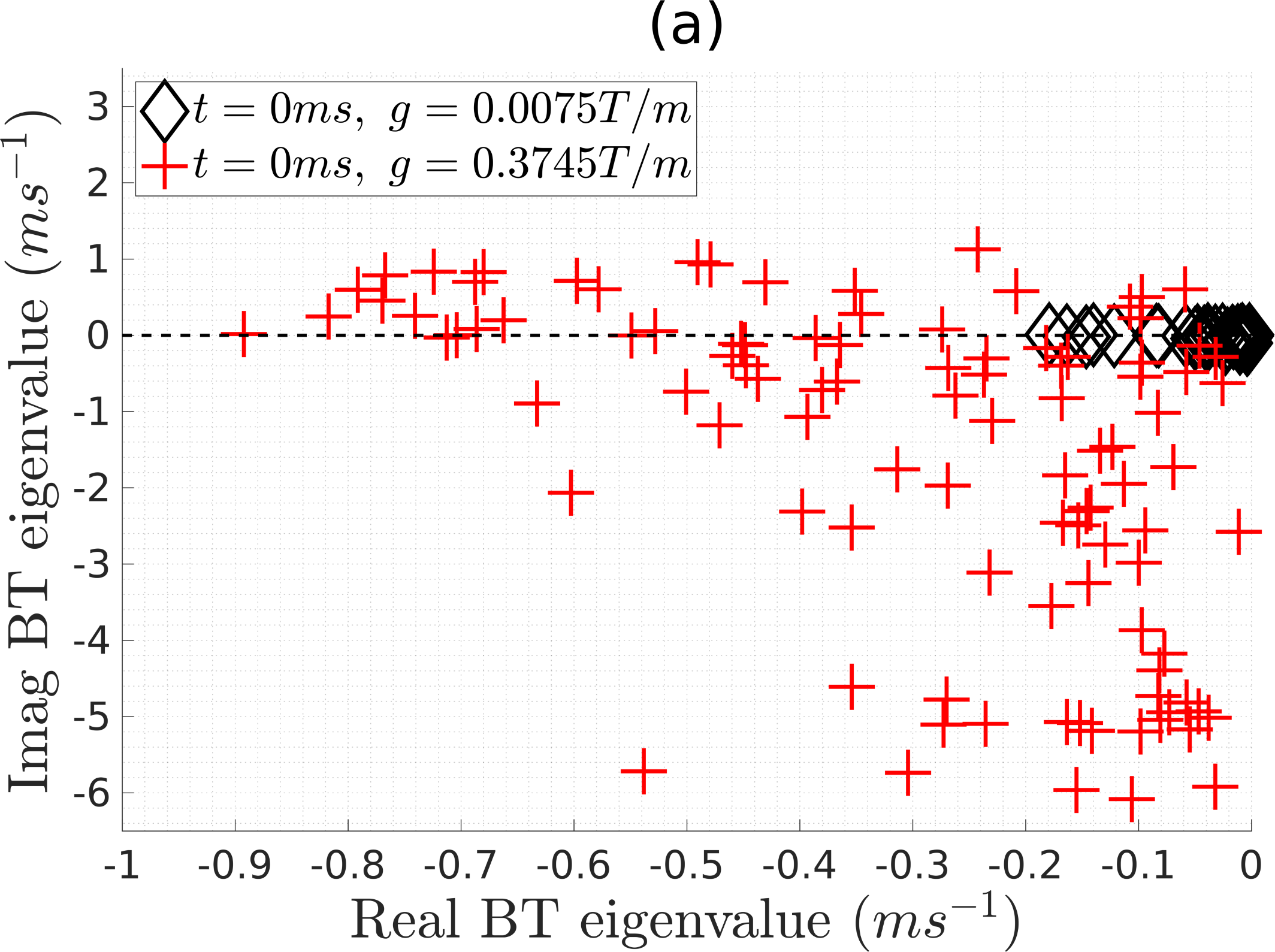}
	\includegraphics[width=0.49\textwidth]{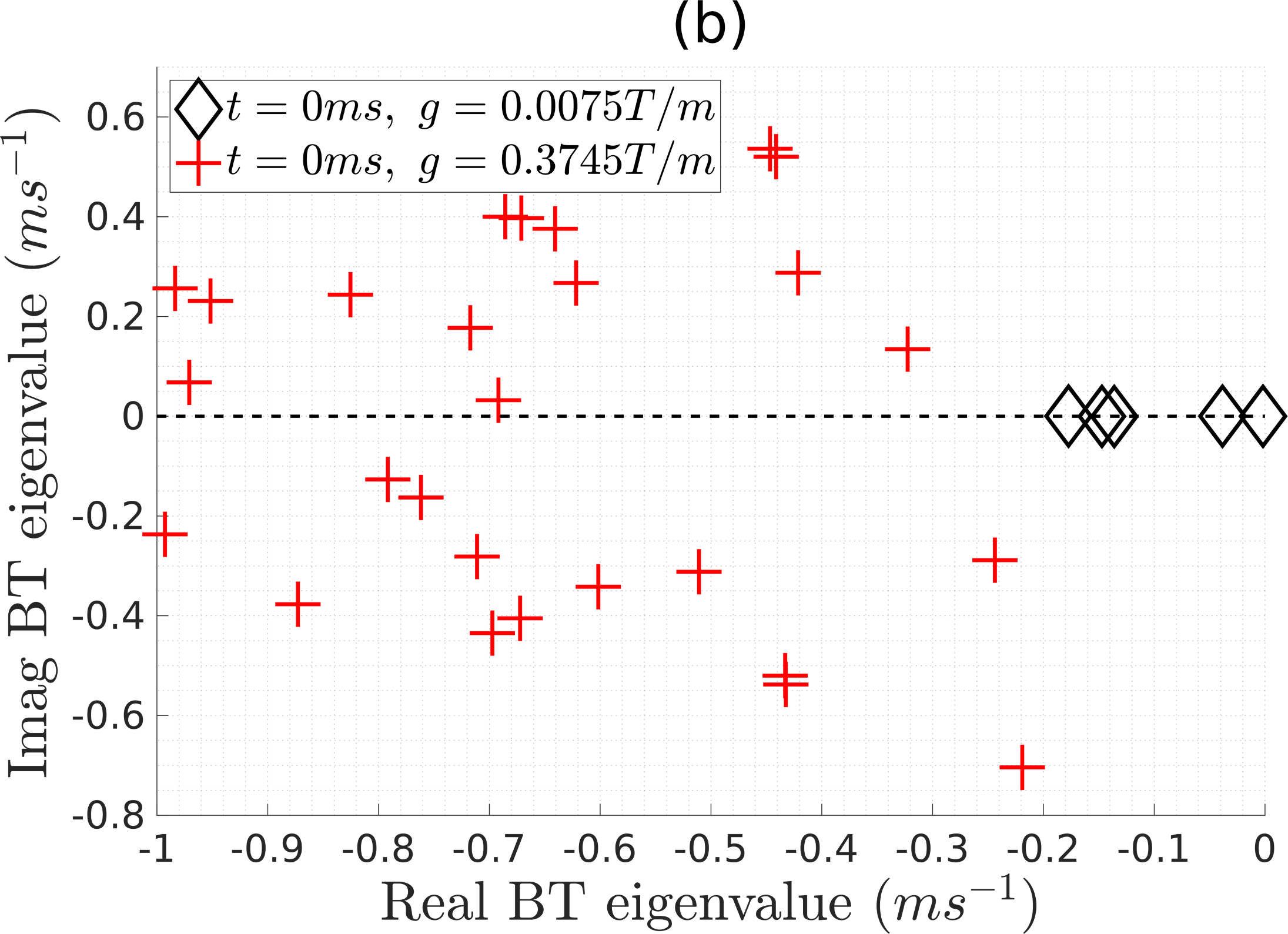}
	\caption{The eigenvalues of the significant Bloch-Torrey eigenfunctions after the projection of the
	initial spin density ($t=0ms$) onto the space of Bloch-Torrey eigenfunctions. 
	Along the x-axis is plotted the real part of the Bloch-Torrey eigenvalues and along the y-axis is plotted the imaginary part of the 
	Bloch-Torrey eigenvalues.
	The gradient direction is $\bug =[-0.3536   -0.6124   -0.7071]$ and two gradient amplitudes, 
	$\vert \bg \vert = 0.0075 \qunit$  and $\vert \bg \vert = 0.3745 \qunit$, were computed.
	(a) the neuron {\it 03b\_spindle4aACC}.  (b) its soma.}
	\label{fig:discretization_parameters_time_BT}
\end{figure}

We projected the initial spin density, which is a constant function, onto the Bloch-Torrey eigenfunctions.
We normalized the initial spin density as well as the Bloch-Torrey eigenfunctions so that they have unit L2 norm (integral of the square of the function over the geometry).
We call a Bloch-Torrey eigenfunction {\it significant} if the projection coefficient is greater than $0.01$ and the real part of 
its eigenvalue is greater than $-1 \tunit^{-1}$.  The latter requirement means we are looking at Bloch-Torrey eigenfunctions that do not 
decay too fast. 
In Figure \ref{fig:discretization_parameters_time_BT} we show the complex eigenvalues of the {\it significant} Bloch-Torrey eigenfunctions for the whole neuron and for the soma. \soutnew{}{Since we projected the initial spin density, this corresponds to $t=0ms$.} We see that at the higher gradient amplitude, the
Bloch-Torrey eigenvalues have a wider range in both their real parts as well as their imaginary parts, 
than at the smaller gradient amplitude.  A larger range of real parts indicates faster
transient dynamics during the two gradient pulses, and a larger range of imaginary parts indicates more 
time oscillations.  Both explain why more time discretization points are
needed at higher $\vert\bg\vert$.  For the whole neuron, at the higher gradient amplitude,
there is a much larger range in the imaginary parts of the Bloch-Torrey eigenvalues than for the soma, 
which is why there are so many more oscillations in the whole neuron magnetization than the soma magnetization in Figure
\ref{fig:magnetization_oscillation}.

To give some indication of the needed space discretization for diffusion MRI simulations, 
we computed the eigenfunctions and the eigenvalues of the Laplace operator, 
which governs the dynamics of the magnetization between the gradient pulses ($t\in[\delta,\Delta]$).   
The Laplace eigenfunctions and eigenvalues on the neuron are numerically computed using the Matrix
Formalism Module \cite{li2019practical} within the SpinDoctor Toolbox. 

\begin{figure}[H]
	\centering
	\includegraphics[width=0.49\textwidth]{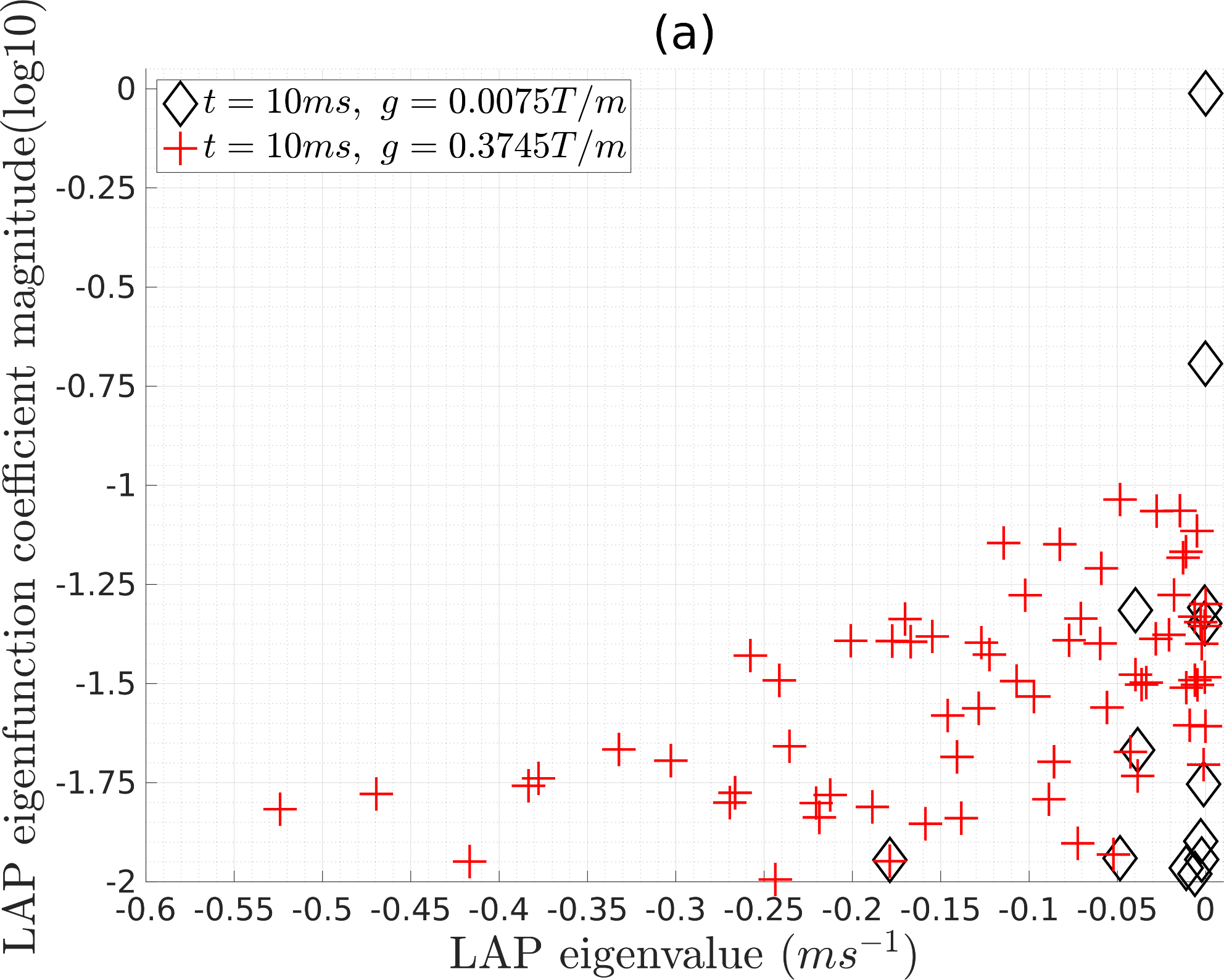}
	\includegraphics[width=0.49\textwidth]{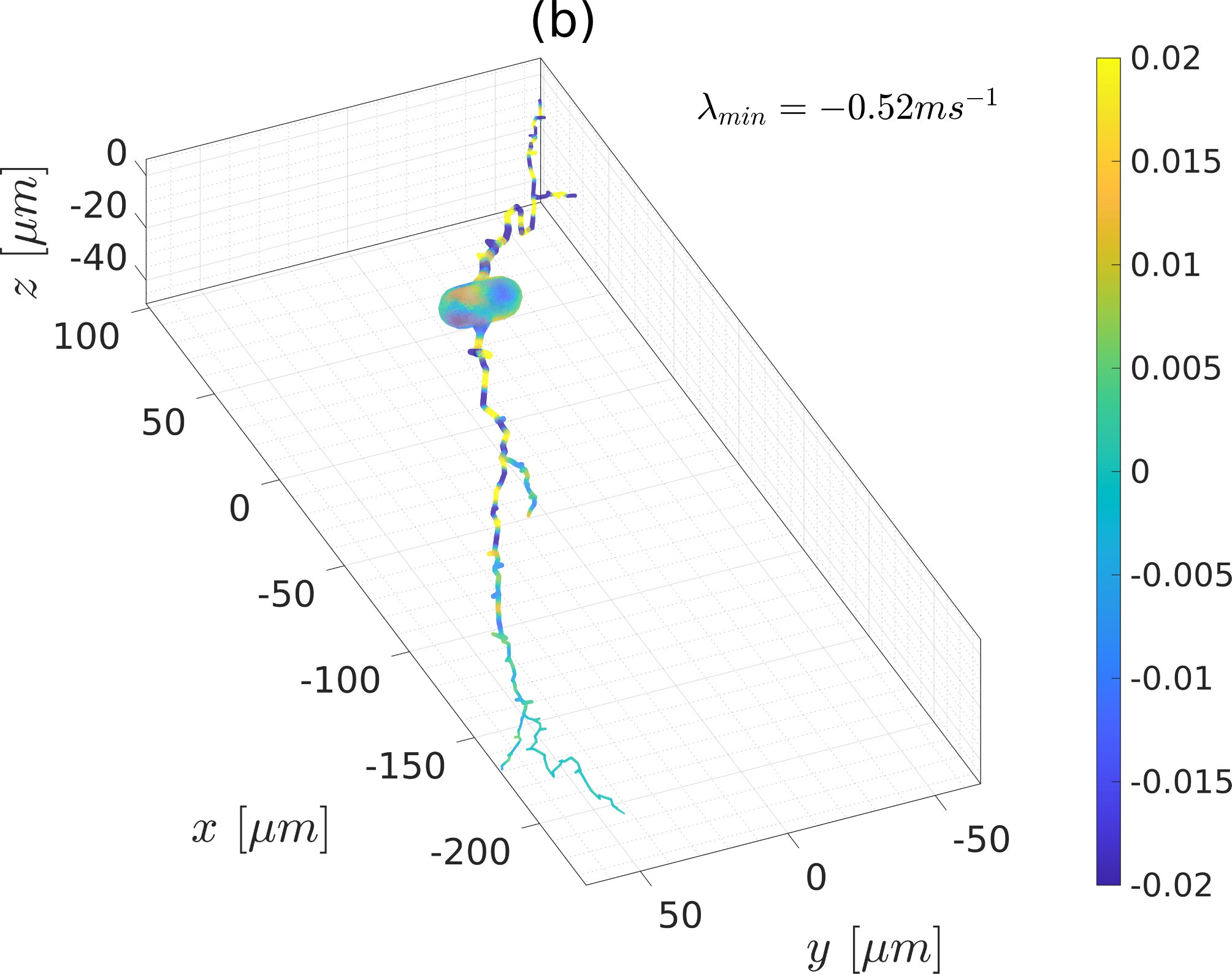}
	\caption{(a) The magnitude of the coefficients of significant Laplace eigenfunctions, after the projection of the magnetization
	at $t=\delta=10\tunit$ unto the space of the Laplace eigenfunctions, plotted against the Laplace eigenvalues, 
	for the whole neuron {\it 03b\_spindle4aACC}.  The experimental parameters are: PGSE ($\delta=10\tunit,\Delta=43\tunit$), gradient direction $\bug =[-0.3536   -0.6124   -0.7071]$.
	Two gradient amplitudes, $\vert \bg \vert = 0.0075 \qunit$  and $\vert \bg \vert = 0.3745 \qunit$ were simulated, equivalent to
	$b = 15.9\bunit$ and $b = 39666.7\bunit$, respectively.  The magnitude of the coefficients are shown in the log 10 scale. 
	(b) The significant Laplace eigenfunction with the most negative eigenvalue.  The color scale is intentionally limited to have a smaller 
	range than the extreme values of the eigenfunction to make the spatial oscillations of the eigenfunction more visible.}
	\label{fig:discretization_parameters_time_LAP}
\end{figure}

We projected the magnetization solution at $t=\delta = 10\tunit$ for 
PGSE ($\delta=10\tunit,\Delta=43\tunit$) for the whole neuron onto the space of the Laplace
eigenfunctions and we show the magnitude of the coefficients of the projections in Figure
\ref{fig:discretization_parameters_time_LAP}.  We see that there are significant
Laplace eigenfunctions with eigenvalues ranging from $\lambda = - 0.52 \tunit^{-1}$ to $0$.
We plot the significant Laplace eigenfunction with the most negative eigenvalue
and we see that this eigenfunction is very oscillatory in space.  To correctly
capture the dynamics of this eigenfunction, it is necessary to have a space discretization that is small
compared to the ``wavelength" of this eigenfunction, which we estimate to be about $10\mu m$
by visual inspection of the space variations shown in Figure \ref{fig:discretization_parameters_time_LAP}.
By choosing $Htetgen = 0.5\mu m$, we are putting about 20 space discretization points per ``wavelength".

\subsection{\soutnew{}{Biomarkers of the soma size}}
\label{sec:biomarkers_soma}
\marginparnew{New section added in revision}
As we have shown in Figure \ref{fig:high_bvalue_fitting}, the linear relationship between $S_{ave}(b)$ and $\frac{1}{\sqrt{b}}$, in other 
words, the power law scaling of the direction-averaged diffusion MR signal \cite{Veraart2019}, doesn't hold due to the presence of the the soma and the exchange effects between the soma and the dendrites. The breakdown of the power law is also observed in \cite{SANDI} and \cite{veraart2020noninvasive}. By leveraging the collection of the realistic neuron meshes, in this section, we statistically show that the deviation from the power law has the potential to serve as biomarkers for revealing the soma size.

In order to do this, we conducted the following simulations that are slightly different than the constant ($\delta$,$\Delta$) experiments in 
\cite{Veraart2019,SANDI,veraart2020noninvasive} and shown in Figure \ref{fig:high_bvalue_fitting}. The signals are numerically computed using the Matrix Formalism Module within the SpinDoctor Toolbox. In the following, we held the gradient amplitude
constant, $\gamma|\bg| = 10^{-5} s^{-1}\cdot mm^{-1}$, and varied $\delta$ to obtain a wide range of b-values, all the while choosing $\Delta=\delta$ (PGSE sequence).  The simulations were conducted in 64 gradient directions 
and the signals were averaged over these directions.  This was performed for the full set of 65 neuron meshes.

\begin{figure}[H]
	\centering
	\includegraphics[width=0.7\textwidth]{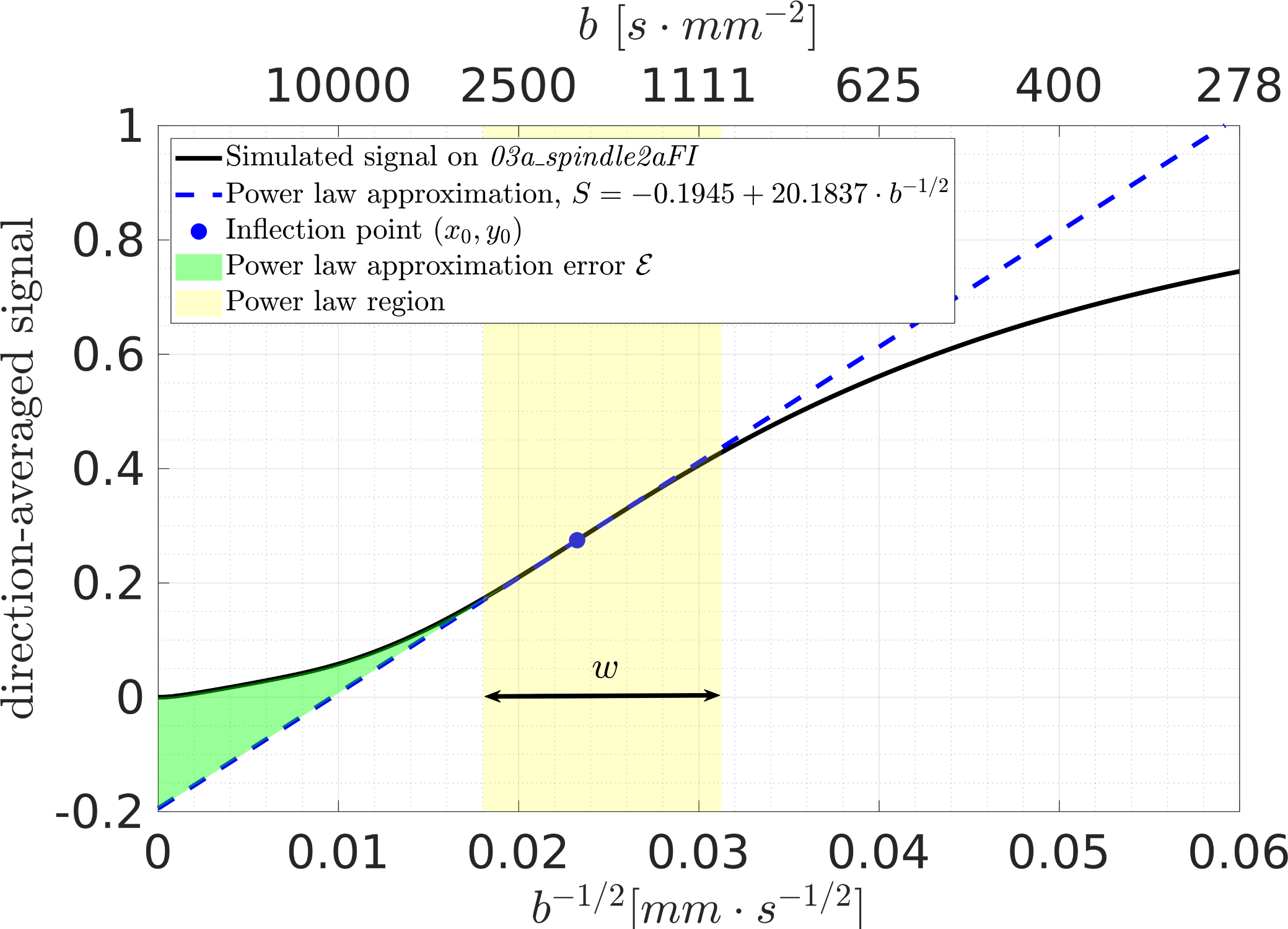}
	\caption{The direction-averaged signal curve for the neuron {\it 03a\_spindle2aFI}. The signals are numerically computed using the Matrix Formalism Module within the SpinDoctor Toolbox. The $S_{ave}(b)$ was averaged over 64 diffusion directions, uniformly distributed in the unit sphere, and it is normalized so that $S_{ave}(b=0)=1$. The b-values are greater than 278 $s/mm^{2}$ and the diffusivity is $\Dintr = 2 \times 10^{-3}\dunit$.  The gradient amplitude is constant, $\gamma |\bg| = 10^{-5} s^{-1}\cdot mm^{-1}$, and $\delta$ was varied to obtain a wide range of b-values, all the while choosing $\Delta=\delta$ (PGSE sequence).
	The blue dot indicates the inflection point of the simulated signal curve. The power law is fitted around the inflection point. The power law region is the width of the range where the relative error between the simulated signal and the power law approximation is less than 2\%. The area between the simulated curve and the power law to the left of the inflection point represents the approximation error of the power law.}
	\label{fig:possible_biomarkers}
\end{figure}

In Figure \ref{fig:possible_biomarkers} we show an example of the simulated signal curve and the power law approximation for the neuron {\it 03a\_spindle2aFI}.  From the direction-averaged simulated signals, we find the inflection point (blue dot) of the signal curve (black curve).  We fit the power law (straight blue dashed line) around the inflection point. The power law region is the range where the relative error between the simulated signal curve and the power law fit is less than 2\% (width of the yellow region) and the approximation error is estimated by the area between the signal curve and the power law fit {\it to the left of the inflection point} (the green area) .

In order to characterize the influence of soma on the power law approximation, we chose the following 6 candidate biomarkers:
\begin{itemize}
	\item $x_0$: the x-coordinate of the inflection point;
	\item $y_0$: the y-coordinate of the inflection point;
	\item $c_0$: the y-intercept of the power law fit;
	\item $c_1$: the slope of the power law fit;
	\item $\mathcal{E}$: the power law approximation error;
	\item $w$: the width of the power law region.
\end{itemize}

A statistical study of the above 6 candidate biomarkers on the collection of the 65 neurons in the Neuron Module was performed. Since the undersampling when $\frac{1}{\sqrt{b}}$ approaches $0$ could produce significant numerical error, we only kept the neurons whose $x_0$ are greater than 0.016 $mm\cdot s^{-1/2}$. In total, 28 spindle neurons and 21 pyramidal neurons were retained.

\begin{figure}[H]
	\centering
	\includegraphics[width=0.98\textwidth]{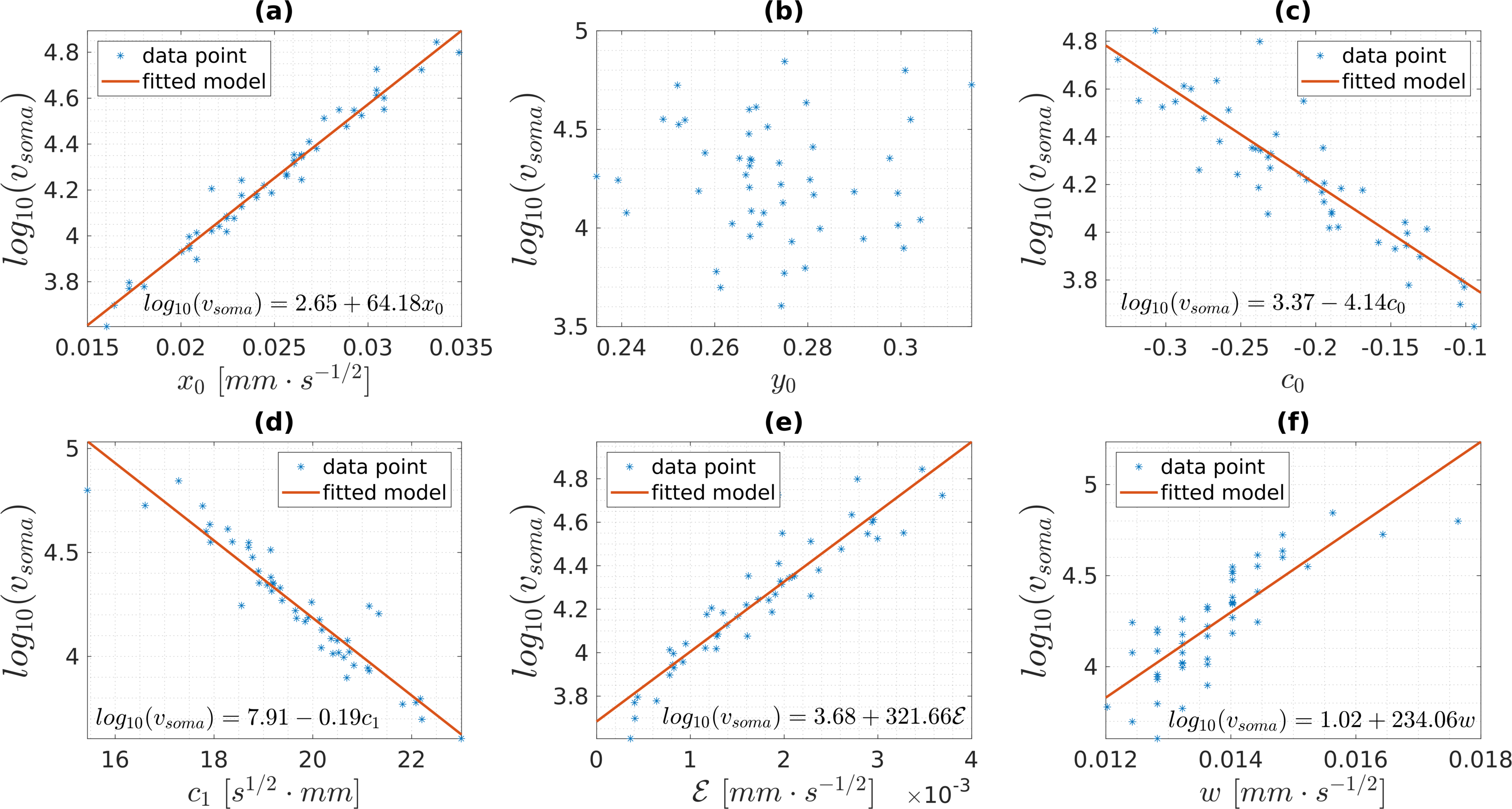}
	\caption{(a) the logarithm of soma volume vs. the x-coordinate of the inflection point $x_0$. (b) the logarithm of soma volume vs. the y-coordinate of the inflection point $y_0$. (c) the logarithm of soma volume vs. the y-intercept of the power law $c_0$. (d) the logarithm of soma volume vs. the slope of the power law $c_1$. (e) the logarithm of soma volume vs. the power law approximation error $\mathcal{E}$. (f) the logarithm of soma volume vs. the width of the power law region $w$.  Each blue dot represents the data from one of the 49 neurons (28 spindle neurons and 21 pyramidal neurons) retained for this study.}
	\label{fig:volume_biomarkers}
\end{figure}

\begin{figure}[H]
	\centering
	\includegraphics[width=0.98\textwidth]{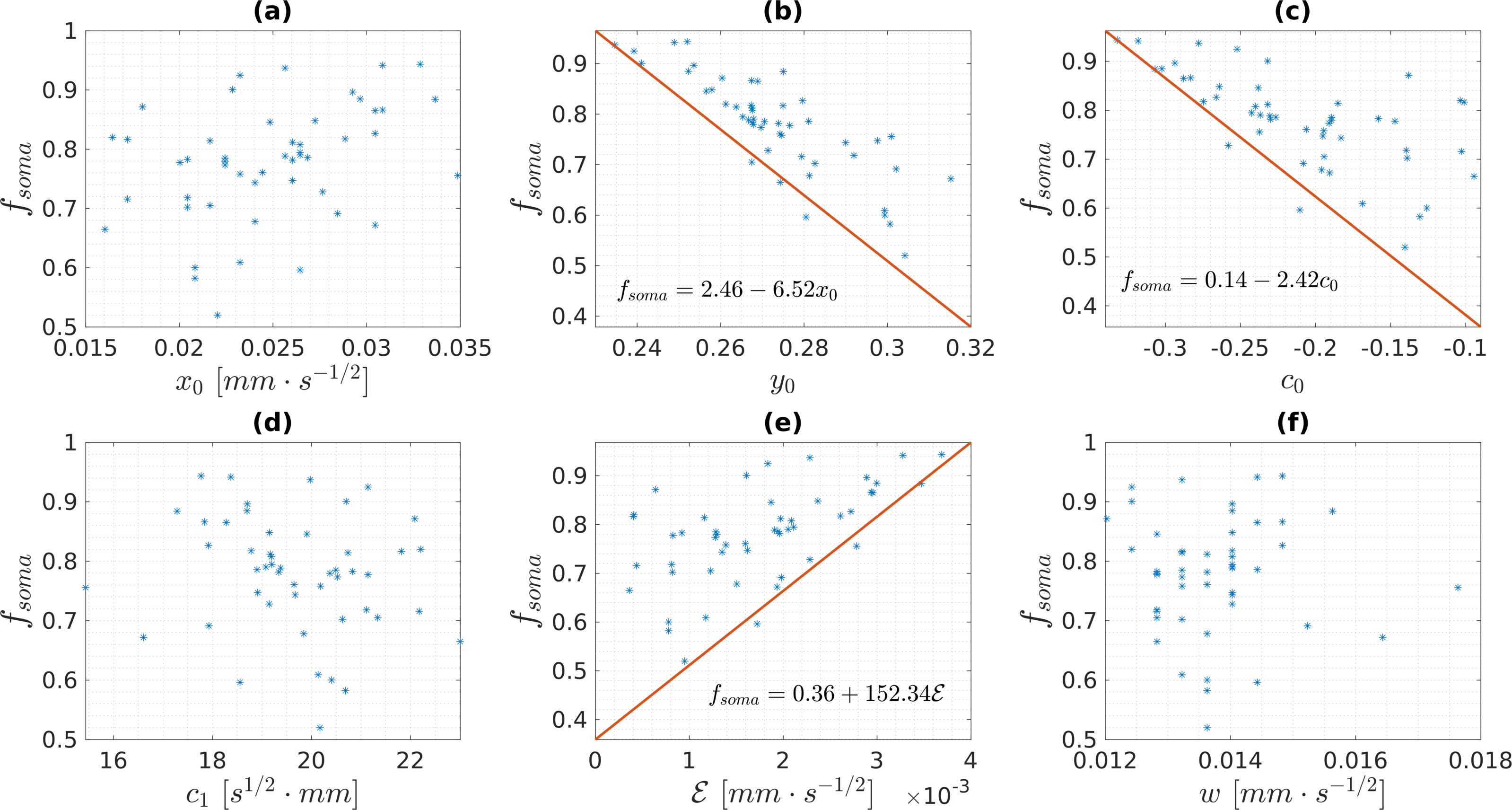}
	\caption{(a) the soma volume fraction vs. the x-coordinate of the inflection point $x_0$. (b) the soma volume fraction vs. the y-coordinate of the inflection point $y_0$. (c) the soma volume fraction vs. the y-intercept of the power law $c_0$. (d) the soma volume fraction vs. the slope of the power law $c_1$. (e) the soma volume fraction vs. the power law approximation error $\mathcal{E}$. (f) the soma volume fraction vs. the width of the power law region $w$.  Each blue dot represents the data from one of the 49 neurons (28 spindle neurons and 21 pyramidal neurons) retained for this study.}
	\label{fig:fraction_biomarkers}
\end{figure}

We first plot the candidate biomarkers with respect to the soma volume $v_{soma}$ in Figure \ref{fig:volume_biomarkers}. Each data point in the figure corresponds to a neuron (for a total of 49).  It can be seen that $x_0$, $c_0$, $c_1$, $\mathcal{E}$, and $w$ exhibit an exponential relationship with the soma volume.  The fitted equations allow us to infer the soma volume by measuring the biomarkers.  We also see
that $y_0$ is not a biomarker for the soma volume.  Similarly, we show the scatter plot of the candidate biomarkers with respect to the soma volume fraction $f_{soma}$ in Figure \ref{fig:fraction_biomarkers}.  In this case, the $x_0$, $c_1$ and $w$ are not biomarkers of the soma volume fraction.  The candidate biomarkers $y_0$, $c_0$ and $\mathcal{E}$ seem capable of indicating the lower bound for the soma volume fraction.

We note that the objective of this section is to give an example of the possible research that can be conducted using the Neuron Module. A more systematic study is needed to get plausible biomarkers for the soma size but this is out of the range of this paper.

\subsection{Additional neuron simulations}
\label{sec:otherneurons}

\begin{figure}[H]
	\centering
	\includegraphics[width=1\textwidth]{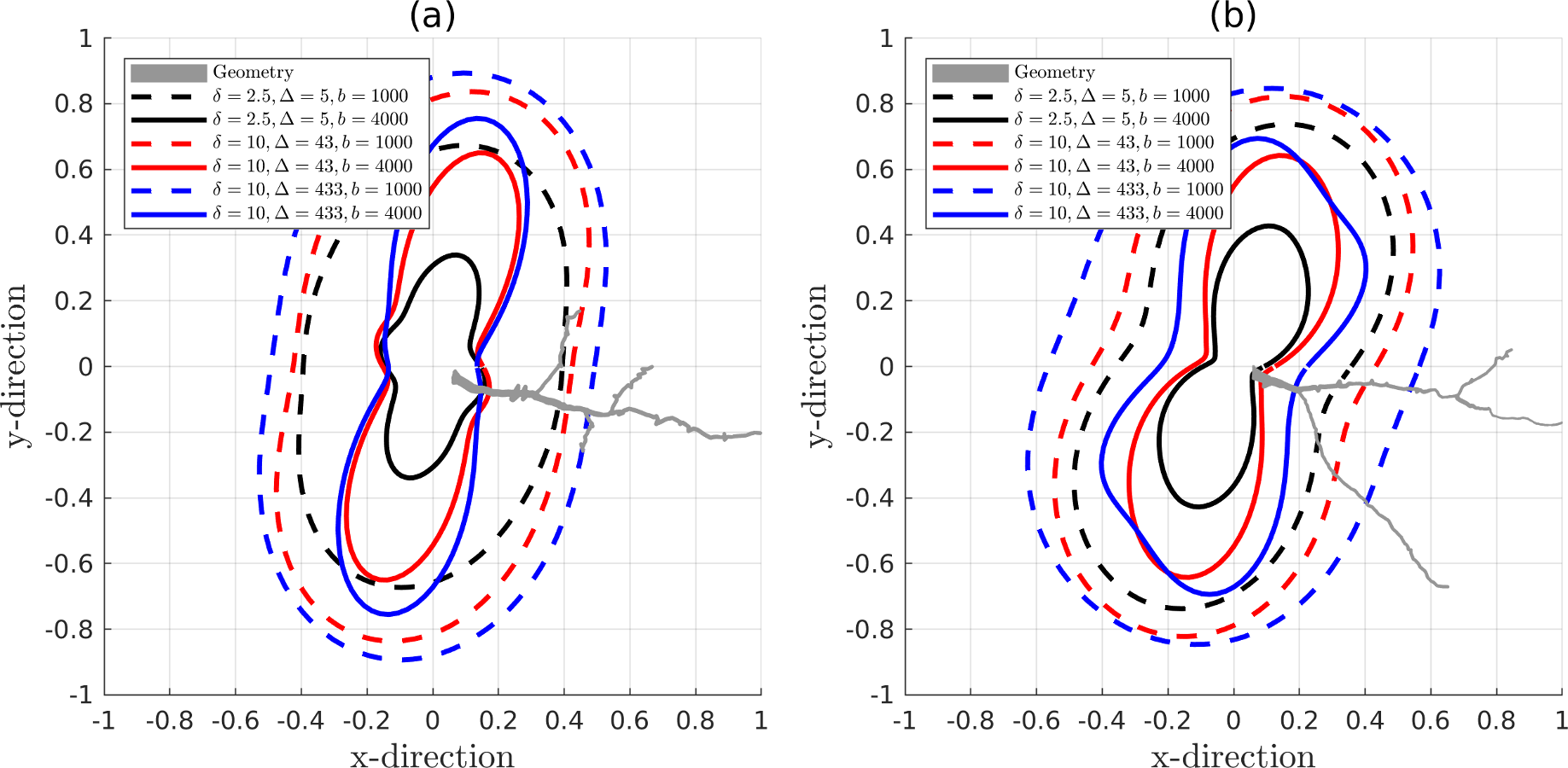}
	\caption{The normalized diffusion MRI signals in 180 directions lying on the $x-y$ plane, uniformly distributed on a unit circle. The distance from each data point to the origin represents the magnitude of the normalized signal which is dimensionless. The simulation parameters are $rtol = 10^{-3}, atol = 10^{-5}, Htetgen = 0.5\mu m$. The diffusion coefficient is $2 \times 10^{-3}\dunit$. (a) one dendrite branch of {\it 04b\_spindle3aFI} (finite elements mesh: 29854 nodes and 95243 elements). (b) one dendrite branch of {\it 03b\_spindle7aACC} (finite elements mesh: 10145 nodes and 28731 elements). \label{fig:hardi_dendrites}}
\end{figure}

Now we show some simulation results on other neuron meshes in our collection.
In Figure \ref{fig:hardi_dendrites}
we compare the diffusion MRI signals due to two different dendrite branches, one from {\it 04b\_spindle3aFI} and one from {\it 03b\_spindle7aACC}.
The first branch has a single main trunk whereas the second branch divides into
two main trunks.  We see at the higher b-value $b=4000\bunit$, at the longest
diffusion time, the signal shape is more elongated (perpendicular to the main
trunk direction) for the first dendrite branch than the second.

% FIGURE 13

\begin{figure}[H]
	\centering
	\includegraphics[width=1\textwidth]{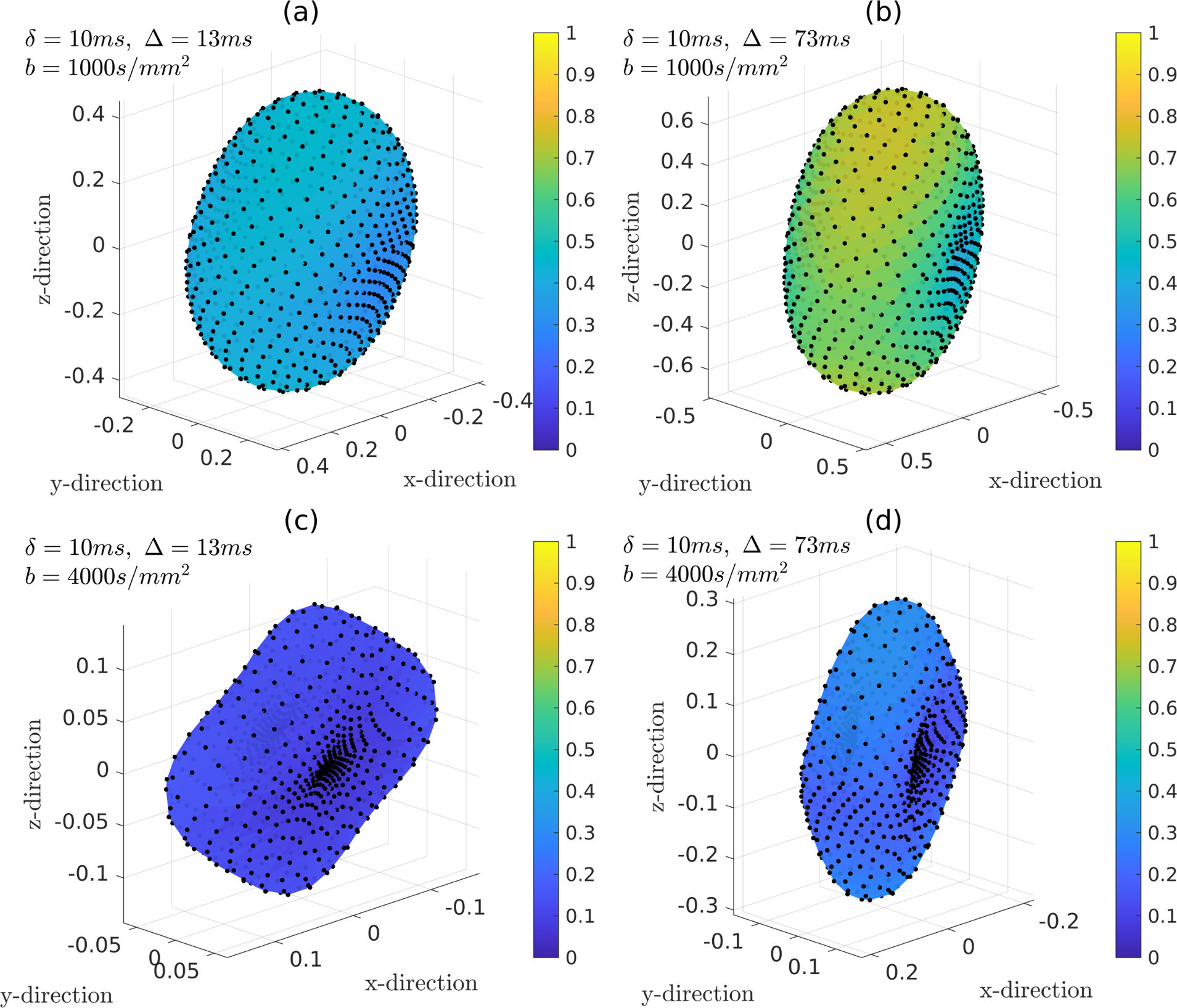}
	
	\caption{The normalized diffusion MRI signals for the neuron {\it 03a\_spindle2aFI} in 720 directions uniformly distributed on a unit sphere.  The color and the distance to the origin of each data point  represent the magnitude of the normalized signal, which is dimensionless. The simulation parameters are $rtol = 10^{-3}, atol = 10^{-5}, Htetgen = 0.5\mu m$.  The diffusion coefficient is $2 \times 10^{-3}\dunit$.
		(a) PGSE ($\delta = 10\tunit$, $\Delta = 13\tunit$), $b = 1000\bunit$. (b) PGSE ($\delta = 10\tunit$, $\Delta = 73\tunit$), $b = 1000\bunit$. (c) PGSE ($\delta = 10\tunit$, $\Delta = 13\tunit$), $b = 4000\bunit$. (d) PGSE ($\delta = 10\tunit$, $\Delta = 73\tunit$), $b = 4000\bunit$.  The number of finite elements nodes and elements for the neuron are $49833$ and $169601$, respectively.
		\label{fig:HARDI_simulations}}
\end{figure}

In Figure \ref{fig:HARDI_simulations} we show 3 dimensional HARDI (High Angular Resolution Diffusion Imaging) 
simulation results of the spindle neuron {\it 03a\_spindle2aFI} (cf. Figure \ref{fig:03a_spindle2aFI_mesh}). We plot in Figure \ref{fig:HARDI_simulations} the normalized signal in 720 directions uniformly distributed in the unit
sphere for $b=1000\bunit$ and $b=4000\bunit$.
We see the normalized signal shape in these 720 directions is ellipsoid at the lower b-value and
the shape becomes more complicated at the larger b-value.
At $b=4000\bunit$, there is more signal attenuation
at the shorter diffusion time than at the higher diffusion time.

% FIGURE 14

\section{Discussion}

In a previous publication \cite{lid}, SpinDoctor, a MATLAB-based diffusion MRI simulation toolbox, was presented.
SpinDoctor allows the easy construction of multiple compartment models of the brain white matter,
with the possibility of coupling water diffusion between the geometrical compartments by permeable membranes.
In \cite{lid}, in addition to white matter simulations,
SpinDoctor's ability to import and use externally generated meshes provided by the user was illustrated
with a neuronal dendrite branch simulation.
This capability is expected to be very useful given the most recent developments in simulating ultra-realistic virtual tissues,
typified by recent work such as \cite{Palombo2019,Ginsburger2019}, which were meant to facilitate Monte-Carlo type simulations.

In order to enrich the publicly available geometrical meshes that can be used for diffusion MRI simulations,
we implemented the Neuron Module inside SpinDoctor.
We have created high quality volume tetrahedral meshes for a group of 36 pyramidal neurons
and a group of 29 spindle neurons. Surface triangulations can be obtained from the volume meshes in a natural way and can be
used for Monte-Carlo simulations.

We cite two very recent works in \cite{Palombo2019,Ginsburger2019} that describe
new algorithms for generating relevant tissue and cell geometries for diffusion MRI simulations.
These two works are similar in spirit to ours, namely, the common idea is to provide synthetic but realistic cell/tissue geometries.
While they use the generated geometries to conduct Monte-Carlo simulations, we principally use
finite elements.  However, in theory, there is nothing preventing conducting either types of simulations on any
high quality surface triangulation.

In terms of the synthetic tissue/cell mesh generation problem, the work in \cite{Ginsburger2019} is more about
the brain white matter.  The work that is closer to ours is \cite{Palombo2019}, which is about
creating 3-dimensional synthetic neurons based on realistic neuron morphology statistics.
That paper contained detailed information about generating synthetic neuron skeletons (tree structure and
branch diameter information), which is analogous to the neuron information in the SWC format available from {\it NeuroMorpho.Org}.
In addition, they used BLENDER and the BLENDER added-on "SWC Mesh" to generate surface triangulations for some neurons
using the neuron information from {\it NeuroMorpho.Org}.
The salient points of our work described in this paper, contrasted with \cite{Palombo2019}, are the following:
\begin{enumerate}
	\item we do not generate synthetic neuron skeletons, we only import existing neuron skeletons from
		      {\it NeuroMorpho.Org};
	\item we tested the BLENDER added-on "SWC Mesh", and we found that we were not able to use it to generate surface triangulations from the neuron skeletons provided by {\it NeuroMorpho.Org} in a simple or automatic way, at least not
	      for our collection of neurons;
	\item following the approach described in this paper, we generated 65 high quality surface triangulation of realistic neurons;
	\item we provide the high quality realistic neuron meshes as a publicly available library;  we provide a User Guide to explain how to use these meshes in diffusion MRI simulations;  this saves the user from having to spend a lot of time to generate surface meshes from neuron skeleton information;
	\item we showed that the neuron and cell component meshes can be used by both finite elements and Monte-Carlo simulations.
	\item the neuron meshes we generated can be coupled with finite elements discretization to compute the eigenfunctions and eigenvalues of the Bloch-Torrey and the Laplace operators;
\end{enumerate}

In addition, in this paper,
\begin{enumerate}
	\item we performed an accuracy and computational timing study of the serial SpinDoctor finite elements simulation and a GPU implementation of the Monte-Carlo simulation; We showed that at equivalent accuracy, if only one gradient direction needs to be simulated, SpinDoctor is faster than GPU Monte-Carlo.  Because the GPU Monte-Carlo method is inherently parallel, if many gradient directions need to be simulated, there is a break-even point when GPU Monte-Carlo becomes faster than SpinDoctor.  In particular,
	      at equivalent accuracy, we showed the ratio between the GPU Monte-Carlo computational time and the SpinDoctor computational time for 1 gradient direction ranges from 31 to 72 for the whole neuron simulations.
	\item we explained the choice of space and time discretization parameters in terms of the eigenvalues and the eigenfunctions of the Laplace and Bloch-Torrey operators of the computational domain in question and illustrated the differences between high gradient amplitude simulations versus low gradient amplitude simulations.  We believe this will help to guide the choice of simulation parameters (number of spins and the time step size) for Monte-Carlo simulations as well.
\end{enumerate}

The HARDI simulations we performed on the neurons confirm some established views of
diffusion measurements \cite{kiselev2017fundamentals}:  1) the diffusion tensor is the prevalent contribution at long diffusion times and low gradient amplitudes, exemplified by the elliptical shape of the signal attenuation;  2) the time dependence of the diffusion tensor can be seen by the fact that the ellipses become larger (more signal, less attenuation, a smaller apparent diffusion coefficient) at longer diffusion
times; 3) at higher gradient amplitudes and shorter diffusion times, the HARDI signal attenuation has a more complicated shape, 
probably due to 
the ``stick'' contributions of the dendrites;  this additional complexity of the diffusion behavior could be explainable with 
higher order diffusion characteristics, in other words, higher order cumulants beyond the diffusion tensor.  A more detailed theoretical analysis of diffusion in neurons would be made in the future using the numerically computed eigenvalues and the eigenfunctions of the Laplace and Bloch-Torrey operators.

\soutnew{}{Taking advantage of the realistic neuron meshes, we were able to illustrate the potential of our work. We achieved this through 3 examples: first, we showed the capability to accelerate diffusion MRI simulations; second, we tested the hypotheses of the compartment-based imaging models; third, we showed how our tool can help to design new imaging methods.}

\soutnew{}{Regarding the first point, we compared the computational times used by SpinDoctor and the GPU Monte-Carlo program thoroughly in Section} \ref{sec:simul_timing} (see Table \ref{tab:computational_time}). \soutnew{}{Our methodology has advantages in speed and accuracy, thereby making our work a promising alternative to Monte-Carlo methods which are widely used in many diffusion MRI studies, e.g.,} \cite{Rensonnet2019, rafael2020robust, jelescu2017design}. \soutnew{}{As for the testing of imaging models, most compartment-based imaging models, e.g.,} \cite{jespersen2007modeling, zhang2012noddi, novikov2018rotationally, kaden2016multi, reisert2017disentangling} \soutnew{}{assume that the brain micro-structure consists of two impermeable compartments, namely intra-neurite space and extra-neurite space. These models succeeded in extracting micro-structure information such as neurite orientation dispersion, axon radii and neurite density in white matter, e.g.,} \cite{christiaens2020need, alexander2019imaging}.  However, some studies \cite{mckinnon2017dependence, Veraart2019, veraart2020noninvasive, palombo2018compartment, henriques2019microscopic, Jespersen2019} \soutnew{}{have shown that their assumptions are invalid in gray matter at large b-values.  Such a conclusion is in accordance with our results shown in Figure \ref{fig:high_bvalue_fitting} and discussed in Section \ref{sec:simul_highb}, illustrating that we are capable of simulating recent experimental
	findings, and according to Table \ref{tab:computational_time}, much faster than previous approaches.}

\soutnew{}{Regarding the use of SpinDoctor to assist the design of new voxel-level models, we take the example recently published by Palombo et al.} \cite{SANDI}. \soutnew{}{In their work, they took the diffusive restriction effect caused by soma into account and proposed a new compartment-based model that can hold in gray matter. To assess the validity regime of the non-exchanging compartment model for different diffusion times and b-values, they simulated simplified neuron models in Camino. Specifically, Palombo et al. compared the simulated ADC by connecting or disconnecting the cylinders and the sphere. In Section} \ref{sec:exchange_between_soma_dendrites}, \soutnew{}{we compared the diffusion MRI signal of a connected neuron and a disconnected neuron, which we show in Figure} \ref{fig:error_compositesignal}. \soutnew{}{Since we have shown that SpinDoctor has advantages in speed and accuracy, studies such as Palombo et al.’s could gain by simulating realistic neurons with less use of computational resources.  Besides helping to design novel compartment-based models, our work also points to new research possibilities that have been previously limited by the moderate efficiency of Monte-Carlo methods, for example, the statistical study performed in Section} \ref{sec:biomarkers_soma}.

In summary, we believe our work can add substantially to the understanding of the imaging of
neuronal micro-structure (neurite density, neurite orientation dispersion, neuronal morphology) \cite{Jespersen2007,zhang2012,Palombo2016,Novikov2018,Lampinen2017,Lampinen2019,Jespersen2010}.
In this paper, we have conducted a detailed numerical study of one neuron, the {\it 03b\_spindle4aACC}, to validate our approach. 
\soutnew{}{We have also shown a preliminary statistical study of the entire collection of neurons.}
For the interested reader, we have included numerical simulations of other neurons and cell components in the Supplementary Material.  Clearly, \soutnew{a statistical study}{further detailed statistical studies} of a large number of neurons is the logical next step to our work.

Our work sets the stage for a systematic study of \soutnew{}{the connection between} the diffusion
MRI signal \soutnew{from different types of neurons}{and neuron morphology} by the diffusion MRI
community (for preliminary results, see \cite{Wassermann2018, Menon662601} \soutnew{}{and Section \ref{sec:biomarkers_soma}}).
We hope this work contributes to further understanding of \soutnew{the}{that} relationship \soutnew{between cell
	morphology and the resulting diffusion MRI signal}{} and aids in better signal model formulation in the future.
If there is sufficient interest from the modeling community, we will add high quality meshes
of other realistic neurons in the future.

As this time, we have not implemented the Neuron Module for coupled compartments linked
by permeable membranes.  Rather, the diffusion MRI signal is computed with zero permeability on the compartment
boundaries.  The current emphasis of the Neuron Module is to show how the geometrical structure
of the neurons affect the diffusion MRI signal.
Thus, some of the input parameters related to multiple compartment models in SpinDoctor are not
applicable in the current version of the Neuron Module.  However, we have kept the exactly same input
file formats in anticipation of the future development of the Neuron Module for
permeable membranes.

In the Supplementary Material, we list the expected input files, as well as the important functions relevant to the Neuron Module.  Sample output figures are also provided there.  The toolbox SpinDoctor and the Neuron Module
as well as the User Guide are publicly available at:
\begin{center}
	\url{https://github.com/jingrebeccali/SpinDoctor}
\end{center}

The complete set of the volume tetrahedral meshes of the whole neurons as well as the corresponding soma and
dendrite branches for the group of 36 pyramidal neurons
and the group of 29 spindle neurons are publicly available at:
\begin{center}
	\url{https://github.com/van-dang/RealNeuronMeshes}
\end{center}
The names and sizes of the finite elements meshes of the 65 neurons and the morphological characteristics of the neurons are listed in the Supplementary Material.

\section{Conclusion}\label{Conclusion}

We presented the Neuron Module that we implemented in the Matlab-based diffusion MRI simulation toolbox SpinDoctor.
We constructed high quality volume tetrahedral meshes
for a group of 36 pyramidal neurons and a group of 29 spindle neurons.
Using the Neuron Module,
the realistic neuron volume tetrahedral meshes can be seamlessly coupled with the functionalities of SpinDoctor
to provide the diffusion MRI signal attributable to spins inside neurons for general diffusion-encoding sequences,
at multiple diffusion-encoding gradient amplitudes and directions.
In addition, we have demonstrated that these neuron meshes can be used to perform Monte-Carlo diffusion MRI simulations as well. We gave guidance in the choice of simulation parameters for both finite elements and Monte-Carlo approaches using the eigenfunctions and eigenvalues of the Bloch-Torrey and Laplace operators that we computed numerically on these neuron meshes with finite elements discretization. \soutnew{}{Finally, we
	performed a statistical study on the collection of neurons in the Neuron Module and tested some candidate biomarkers that can potentially reveal the soma size.  We hope this study can inspire new imaging methods in the future.}

\section*{Acknowledgment}
Van-Dang Nguyen was supported by the Swedish Energy Agency with the project ID P40435-1 and MSO4SC, the grant number 731063.  We would like to thank ANSA from Beta-CAE Systems S. A., for providing an academic license. The authors are extremely grateful to Dr. Khieu Van Nguyen for his generous help in explaining and modifying his GPU-based Monte-Carlo code, making it possible for us to conduct the timing and accuracy comparison study in this paper.
% \section*{Appendix}



%\appendix

%% References

\newpage
\bibliographystyle{elsarticle-num}
%\bibliography{myref_VDN}
\bibliography{myref_final}





% \end{thebibliography}

\end{document}

%% file: my_packages.tex
\usepackage[utf8]{inputenc}
\usepackage{graphicx,amssymb,amstext,amsmath}
\usepackage{amsmath,amsxtra,amssymb,latexsym, amscd,amsthm}
\usepackage{indentfirst}
\usepackage[mathscr]{eucal}
\usepackage{graphicx}
\usepackage{ulem}
\usepackage{enumitem}
\usepackage{url}
\usepackage{subfig}
\usepackage{bm}
\usepackage{multirow}
\usepackage{algorithm2e}
\RestyleAlgo{boxed}
\RestyleAlgo{boxruled}
\usepackage{tikz}
\usetikzlibrary{matrix,shapes,arrows,positioning,chains}
\tikzset{
block/.style={
    rectangle,
    draw,
    text width=17em,
    text centered,
},
decision/.style={
    rectangle,
    draw,
    text width=17em,
    text centered,
    rounded corners
},
cloud/.style={
    draw,
    ellipse,
    minimum height=2em
},
descr/.style={
    fill=white,
    inner sep=2.5pt
},
connector/.style={
    -latex,
    font=\scriptsize
},
rectangle connector/.style={
    connector,
    to path={(\tikztostart) -- ++(#1,0pt) \tikztonodes |- (\tikztotarget) },
    pos=0.5
},
rectangle connector/.default=-2cm,
straight connector/.style={
    connector,
    to path=--(\tikztotarget) \tikztonodes
}
}

\usepackage{float}

\usepackage{mathrsfs}
\usepackage{geometry}
\usepackage{fancyhdr}

\usepackage{listings}             % Include the listings-package
\lstdefinestyle{myCustomMatlabStyle}{
  language=Matlab,
  numbers=left,
  stepnumber=1,
  numbersep=10pt,
  tabsize=4,
  showspaces=false,
  showstringspaces=false
}

%% file: my_macros_math.tex
\newcommand{\bx}{\ensuremath{{\bm{x}}}}

\newcommand{\bg}{\ensuremath{{\bm{g}}}}
\newcommand{\dunit}{\,\ensuremath{\text{mm}^2/\text{s}}}
\newcommand{\bunit}{\,\ensuremath{\text{s/mm}^2}}
\newcommand{\tunit}{\ensuremath{\text{ms}}}

\newcommand{\lunit}{\ensuremath{\mu\text{m}}}
\newcommand{\qunit}{\ensuremath{\text{T/m}}}

\newcommand{\bn}{\ensuremath{{\bm{n}}}}

\newcommand{\ben}{\begin{equation*}}
\newcommand{\een}{\end{equation*}}
\newcommand{\be} [1] {\begin{equation} \label{#1}}
\newcommand{\ee}{\end{equation}}

\newcommand{\ignore}[1]{}

\newcommand{\Dintr}{\mathcal{D}_0}
\newcommand{\bug}{\ensuremath{{\bm{u_g}}}}

%% file: my_macros_format.tex
\newcommand{\soutnew}[2]{#2}

\newcommand{\marginparnew}[1]{}